\documentclass[12pt,a4paper,final]{iopart}

\pdfoutput=1
\usepackage{amsmath,amssymb,amsfonts}

\usepackage{graphicx}
\usepackage{color}
\usepackage{cite}
\usepackage[breaklinks=true,colorlinks=true,linkcolor=blue,urlcolor=blue,citecolor=blue]{hyperref}
\hypersetup{breaklinks=true,colorlinks=true,citecolor=violet,urlcolor=blue!60!black,linkcolor=red}

\usepackage{braket}
\usepackage{upgreek}
\usepackage{pgf,tikz}
\usetikzlibrary{shapes,calc,decorations.markings}

\def\epc{\, ,}
\def\epp{\, .}
\def\Det{\rm Det}

\begin{document}

\title[Universal terms in the overlap of the XXZ ground state with the N\'eel state]{Universal terms in the overlap of the ground state of the spin-1/2 XXZ chain with the N\'eel state}

\author{Michael Brockmann$^{1,2}$, Jean-Marie St{\'e}phan$^{1,3}$}

\address{
$^{1}$Max-Planck-Institut f\"ur Physik komplexer Systeme,\\ N\"othnitzer Str.~38, 01187 Dresden, Germany\\
$^2$Fachbereich Physik,\, Bergische Universit\"at Wuppertal,\\ Gau{\ss}str.~20, 42119 Wuppertal, Germany\\
$^3$Univ Lyon, Universit\'e Claude Bernard Lyon 1, CNRS UMR 5208, Institut Camille Jordan, 43 blvd. du 11 novembre 1918, F-69622 Villeurbanne cedex, France
}


\date{\today}

\begin{abstract}
We analyze universal terms that appear in the large system size scaling of 
the overlap between the N\'eel state and the ground state of the spin-1/2 XXZ chain in 
the antiferromagnetic regime. In a critical theory, the order one term of the asymptotics 
of such an overlap may be expressed in terms of $g$-factors, known in the context of 
boundary conformal field theory. In particular, for the XXZ model in its gapless 
phase, this term provides access to the Luttinger parameter. In its gapped phase, on 
the other hand, the order one term simply reflects the symmetry broken nature of the 
phase. In order to study the large system size scaling of this overlap analytically and to 
compute the order one term exactly, we use a recently derived finite-size determinant 
formula and perform an asymptotic expansion. Our analysis confirms the predictions 
of boundary conformal field theory and enables us to determine the exponent of the 
leading finite-size correction.
\end{abstract}

\maketitle

\section{Introduction}\label{sec:introduction}
 
Recent years have witnessed renewed interest in the study of quantum phase transi\-tions and quantum critical points in low-dimensional systems at zero temperature. Our theoretical understanding of such points relies mainly on two different but complementary approaches. One is field theory, which captures the important low-energy degrees of freedom that emerge at criticality. Another one is a more direct approach, starting from a concrete microscopic model ---typically a system with $L$ locally coupled elementary degrees of freedom (e.g.~spins or particles)--- and studying the thermodynamic limit (TDL) $L\to \infty$. 

The case of one dimension (1d) is special as very powerful techniques are available on both sides. Most of the critical points in 1d can be described by conformal field theories (CFT) \cite{BPZ,Yellowpages} which are exactly solvable in the sense that it is in principle possible to determine all correlation functions. On the side of the microscopic approach, powerful numerical techniques such as the density matrix renormalization group (DMRG) algorithm \cite{DKSV04,FeWh05,Schollwoeck05,SiKl05} or exact diagonalization \cite{Fehske_book,Weisse2013} are available. An important class of microscopic systems are the so-called Bethe ansatz integrable models. For such models an exact determination of all energy eigenstates is possible \cite{Bethe,GaudinCaux}. Also some correlation functions can be accessed analytically, and the results can be compared with CFT.  

Both sets of techniques have different advantages. One obvious limitation of CFT is that it can, by definition, only access universal numbers that describe the corresponding universality class. So, it is insensitive to the exact details of the underlying microscopic model. Integrability methods based on Bethe ansatz, however, allow to access both, universal and non-universal numbers. In that sense, they provide more information. On the other hand, CFT can describe the low-energy physics of non-integrable models. Said differently, the low-energy effective field theory can still be exactly solvable even though the underlying microscopic model is not. Connecting the two sets of methods is of paramount importance and has been a subject of tremendous effort over the last few years \cite{JimboMiwaetal1993,JimboMiwa,QISMbook,LukyanovTerras,Lyon2009, GKS04a,BDGKSW08}.

One key model where such a connection can be successfully performed is the anisotropic spin-1/2 Heisenberg chain (XXZ model) whose Hamiltonian reads
\begin{equation}\label{eq:xxzham}
  H_L \;=\; \sum_{j=1}^L\left(\sigma_j^x \sigma_{j+1}^x + \sigma_j^y \sigma_{j+1}^y + \Delta \sigma_{j}^z \sigma_{j+1}^z\right)\epc \qquad \Delta\,=\,\cos \gamma\epc \qquad 0\leq \gamma < \pi\epp
\end{equation}
The operators $\sigma_j^\alpha$ act only non-trivially on lattice site $j$ as Pauli matrices $\sigma^\alpha$, $\alpha=x,y,z$, and satisfy periodic boundary conditions, $\sigma_{L+1}^\alpha = \sigma_1^\alpha$. For simplicity, we assume throughout the paper that the system size $L$ is a multiple of four. The parameter $\Delta$ is called the anisotropy of the model. Here, it is restricted to the interval $-1<\Delta\leq 1$, where the XXZ model becomes gapless in the TDL. Later, in Sec.~\ref{sec:Asymptotics_from_BA}, we will also consider the gapped phase $\Delta>1$, where we parameterize the anisotropy by $\Delta=\cosh\eta$, $\eta>0$.

The low-energy behavior of the gapless model \eqref{eq:xxzham} can be described by a free compact boson CFT, also known as Tomonaga-Luttinger liquid \cite{Tomonaga,Luttinger,Giamarchi}, or Gaussian free field in the mathematical literature \cite{Sheffield}. Much effort has also been devoted to the calculation of correlation functions and their large-distance asymptotics, which is a challenging problem \cite{JimboMiwaetal1993,JimboMiwa,QISMbook,LukyanovTerras,Lyon2009}. This has allowed to confirm many predictions of the Tomonaga-Luttinger liquid theory in this system. In particular, one can extract the Luttinger parameter $K$ from the low-energy spectrum or, similarly, from the exponents governing the algebraic decay of bulk correlation functions. It depends on the anisotropy $\Delta=\cos \gamma$ as follows,
\begin{equation}\label{eq:K}
  K \;=\; \frac{\pi}{2(\pi-\gamma)}\epp
\end{equation}
The knowledge of $K$ uniquely determines the underlying CFT. 

In this paper, we demonstrate an alternative way to access the Luttinger parameter, namely, by making a connection to \emph{boundary} CFT, a subject which has been largely developed by Cardy \cite{Cardysurface,Cardy_bccoperatorcontent,CardyVerlinde}. 
We consider the overlap $\mathcal{O}_L$ of the ground state $\ket{\Psi_L}$ of the XXZ model \eqref{eq:xxzham} with the N\'eel state $\ket{N}=\ket{\uparrow\downarrow}^{\otimes L/2}$, 
\begin{equation}\label{eq:OL_def}
  \mathcal{O}_L \;=\; \left| \Braket{N|\Psi_L}\right|^2\epp
\end{equation}
As was shown in Refs.~\cite{Stephan2009} and~\cite{Stephan2011}, its large system size scaling can be analyzed by means of boundary CFT. The overlap may be interpreted as a ratio of partition functions of 2d classical systems (see Sec.~\ref{sec:overlap_CFT}) with boundary conditions that can be argued to renormalize to conformal invariant boundary conditions\cite{Cardy_bccoperatorcontent}. The order one contribution to the overlap is determined by the ``universal ground state degeneracy'', or ``the boundary $g$-factor'' \cite{AffleckLudwig}. It can be viewed as an analogue of the central charge in boundary critical phenomena. In particular, it decreases under RG flow. 

Crucially, the overlap with the N\'eel state can also be evaluated in closed form by means of certain determinant formulas\cite{Tsuchiya, KozlowskiPozsgay, Pozsgay, XXZOverlap, XXZOverlap_Odd} whose derivation is based on algebraic Bethe ansatz. This works, in principle, for all eigenstates of the XXZ model, not only for its ground state. Overlaps of XXZ eigenstates with the N\'eel state were subject of intense study, as they proved instrumental in the understanding of thermalization after quantum quenches \cite{Pozsgay,Wouters_QA,Pozsgay_QA,Brockmann_QA}. We refer the interested reader to the review article \cite{QA_CauxReview}. In the following, however, we are interested in the overlap with the ground state only and, in particular, in its large system size scaling. For this analysis, we will use the determinant formula that is analytically most tractable, and which was derived in Ref.~\cite{XXZOverlap}. 

To be more specific, our main result is the following formula, valid in the `planar regime' $-1 < \Delta < 1$ and obtained by a careful asymptotic analysis of the exact finite-size determinant formula (see Sec.~\ref{sec:Asymptotics_from_BA}),
\begin{equation}\label{eq:OL_result_planar}
  \mathcal{O}_L \;=\; \sqrt{K}e^{-\alpha L}\Big[1\;+\; AL^{-\delta}\;+\;\ldots\Big]\epp
\end{equation}
Here, $K$ is the Luttinger parameter of Eq.~\eqref{eq:K}. The exponential factor $e^{-\alpha L}$ is called the `exponential decay' of the overlap with `decay rate' $\alpha$. Furthermore, we shall refer to the prefactor (here $\sqrt{K}$) as the `order one term', since it determines the $\mathcal{O}(1)$ contribution of $\ln(\mathcal{O}_L)$, as well as to $A$ and $\delta$ as the `amplitude' and the `exponent' of the leading finite-size correction. The dots in Eq.~\eqref{eq:OL_result_planar} represent higher subleading terms. All coefficients are non-trivial functions of $\Delta$. The order one term, which is in exact agreement with the prediction of boundary CFT (see Eq.~\eqref{eq:cftresult} in Sec.~\ref{sec:overlap_CFT}), as well as the exponent $\delta=\min\{1,4K-2\}$ are universal whereas the decay rate $\alpha$, which has the dimension of an inverse length, and the amplitude $A$, which has the dimension length to the $\delta$, are  not. We nevertheless derive a fully explicit integral formula for $\alpha$ and compute the amplitude $A$ (for several values of $\Delta$) to high numerical accuracy (see Sec.~\ref{sec:corrections}). In this sense, we are able to extract even the leading finite-size correction $A/L^{\delta}$ to the overlap, which is not accessible by CFT per se (even though the exponent $\delta$ may be obtained by a perturbed CFT treatment, see Refs.~\cite{perturbedcft_ee1,perturbedcft_ee2,perturbedcft_corners} for such kind of studies). 
 
We further provide results for the `isotropic point' $\Delta=1$ as well as for the `axial regime' $\Delta>1$ (see Sec.~\ref{sec:Asymptotics_from_BA}). The isotropic point can be accessed by taking the limit $\gamma\to 0$ (i.e.~$K\to 1/2$ $\Rightarrow$ ``$\delta\to 0$''), in which Eq.~\eqref{eq:OL_result_planar} turns into
\begin{equation}\label{eq:OL_result_isotropic}
  \mathcal{O}_L\;=\;\frac{1}{\sqrt{2}}e^{-\alpha L}\Big[1\;+\;\mathcal{O}\big(\ln^{-1}(L)\big)\Big]\epp
\end{equation}
The main difference to the planar regime is that the leading finite-size correction is no longer algebraic ($\sim 1/L^{\delta}$) but logarithmic ($\sim 1/\ln L $). 

The result for the overlap in the axial regime reads 
\begin{equation}\label{eq:OL_result_axial}
  \mathcal{O}_L\;=\;\frac{1}{2}e^{-\alpha L}\Big[1\;+\;\mathcal{O}\big(e^{-cL}\big)\Big]\epp
\end{equation}
This formula looks a bit simpler than the result \eqref{eq:OL_result_planar} of the planar regime, as expected for a gapped phase. Note that the finite-size corrections in the axial regime are exponentially small, and that the order one term is independent of the anisotropy $\Delta>1$. The value of the latter can be intuitively under\-stood by considering the limit $\Delta\to\infty$, where the ground state is given by $\ket{\Psi}_L=(\ket{\uparrow\downarrow}^{\otimes L/2}+\ket{\downarrow\uparrow}^{\otimes L/2})/\sqrt{2}$. Hence, the overlap is $\mathcal{O}_L=1/2$ (implying that $\alpha=0$ and $c=\infty$). For $\Delta > 1$ but finite, $\ket{\Psi}_L$ is of course more complicated, but the order one term still reflects the symmetry broken nature of the ground state.

\section{Overlap in boundary conformal field theory}\label{sec:overlap_CFT}

\begin{figure}[tbp]
  \begin{center}
  \includegraphics[width=0.66\columnwidth]{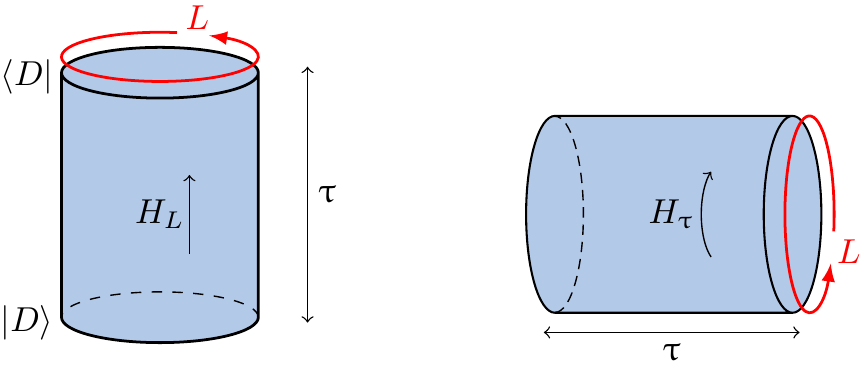}
  \caption{The computation of the partition function of a cylinder can be performed in two equivalent ways, either with a periodic system of size $L$ evolving in imaginary time $\uptau$ with Dirichlet boundary conditions at the top and at the bottom, $Z_{\rm cyl}(L,\uptau)=\bra{D}e^{-\uptau H_L}\ket{D}$ (see left), or with a system of size $\uptau$ with open boundary conditions at inverse temperature $L$, $Z_{\rm cyl}(L,\uptau)={\rm Tr}\{e^{-L H_\uptau}\}$ (see right).}
  \label{fig:cylinders}
  \end{center}
\end{figure}

We focus in this section on the XXZ model in its gapless phase $-1 < \Delta \leq 1$, described by the Hamiltonian \eqref{eq:xxzham}. The ground state of the XXZ model can be seen as the result of an infinite imaginary-time evolution starting from an arbitrary state $\ket{a}$, provided it has nonzero overlap with the ground state,
\begin{equation}
  e^{-\uptau H_L}\ket{a} \;\underset{\uptau\to \infty}{\sim}\; e^{-\uptau E_0}\ket{\Psi_L}\braket{\Psi_L|a}\epp
\end{equation}
Choosing the N\'eel state $\ket{a}=\ket{N}$, for convenience, and using the previous equation, the overlap can be recast as
\begin{equation}
  \mathcal{O}_L \;=\; \lim_{\uptau \to \infty} \frac{\braket{N|e^{-\uptau H_L}|N}\braket{N|e^{-\uptau H_L}|N}}{\braket{N|e^{-2\uptau H_L}|N}}\epp
\end{equation}
This can be interpreted as a ratio of partition functions of two-dimensional classical systems,
\begin{equation}\label{eq:overlapincft}
  \mathcal{O}_L \;=\; \lim_{\uptau \to \infty} \frac{\left[Z_{\rm cyl}(L,\uptau)\right]^2}{Z_{\rm cyl}(L,2\uptau)} \epp
\end{equation}
Here, $Z_{\rm cyl}(L,\uptau)$ denotes the partition function of a cylinder of circumference $L$ and height $\uptau$ (see Fig.~\ref{fig:cylinders}). The boundary conditions at the top and at the bottom are set by the N\'eel state. Now, we make the crucial assumption that the N\'eel boundary condition renormalizes to a conformal invariant boundary condition \cite{Cardysurface, CardyVerlinde}. This argument is very general and only requires to determine the appropriate conformal boundary condition in the continuum limit, if it exists. It was already exploited in Ref.~\cite{Stephan2011}, in order to compute the ratio \eqref{eq:overlapincft} by means of boundary CFT. It can be applied to other models corresponding to different universality classes as well \cite{Stephan2009, Stephan2010,Jacobsen}. Generalizations to chains with open boundary conditions and with universal power-law corrections to the partition function \cite{CardyPeschel} are also possible \cite{Stephan2011,ZBM,Bondesan_rectangle1, Bondesan_rectangle2}.

The conformal boundary condition can be determined as follows. First, we map spin configurations onto discrete height configurations $h_j$, where the height picks up an additive factor, say, $+1$ when encountering an up spin and $-1$ when encountering a down spin, $h_j=\sum_{i=1}^{j} \sigma_i^z$. Thus, for the N\'eel state, we obtain $h_j=(1+(-1)^{j+1})/2$. In the continuum limit, this discrete `height field' $h_j$ is expected to renormalize to a free compact bosonic field, through a procedure known as  bosonization \cite{Giamarchi}. Since the discrete height configuration of the N\'eel state is flat on average, it is natural to expect that the field be constant at the boundary that is set by the N\'eel state. Such a conformal invariant boundary condition is called Dirichlet boundary condition. 

Let us now come back to the computation of the overlap \eqref{eq:overlapincft}. We are left with partition functions of a `free compact boson CFT' on a cylinder with Dirichlet boundary condition, which are well-known \cite{Yellowpages}. We find it instructive to reproduce the derivation of Ref.~\cite{EggertAffleck}, as it allows to make contact with other works which are similar in spirit to ours. Instead of looking at a periodic system which is let evolve in imaginary time $\uptau$, it is equivalent to interpret the CFT partition function as a finite temperature average of a system of width $\uptau$ with free spins at the boundary and a periodic time direction, where the time is now represented by the system size $L$. This is illustrated in Fig.~\ref{fig:cylinders}.

The low-energy physics of an open XXZ chain of length $\uptau$ is well-described by the bosonized form of the XXZ Hamiltonian \cite{EggertAffleck},
\begin{equation}\label{eq:xxzopenbos}
  H_\uptau \;=\; E\uptau \,+\, e \,+\, \frac{\pi v}{\uptau}\left[
  \frac{(S^z)^2}{2K} \,+\, \sum_{k=1}^{\infty} k\left(a_k^\dag a_k+\frac{1}{2}\right)
  \right]\epp
\end{equation}
The coefficients $E$ and $v$  can be interpreted as the extensive part of the ground state energy and as a velocity, respectively. The order one contribution $e$ originates from the boundaries solely. The total magnetization $S^z$ of the chain is an integer, and the $a_k,a_k^\dag$ are bosonic operators that satisfy the Heisenberg commutation relations $[a_k,a_l^\dag]=\delta_{kl}$ and $[a_k^\dag,a_l^\dag]=[a_k,a_l]=0$. It is also understood that $\sum_{k=1}^\infty k=\zeta(-1)=-1/12$ (Zeta regularization). The partition function we are looking for can be written as
\begin{equation}\label{eq:Z}
  Z_{\rm cyl}(L,\uptau) \;=\; {\rm Tr}\big\{e^{-LH_\uptau}\big\}\epc
\end{equation}
where $L$ now plays the role of an inverse temperature. The trace has to be performed over all bosonic modes and all magnetization sectors. Inserting the Hamiltonian \eqref{eq:xxzopenbos} into this definition yields
\begin{equation}\label{eq:cftZ1}
  Z_{\rm cyl}(L,\uptau) \;=\; \frac{e^{-E L \uptau-eL+\pi v L/(24\uptau)}}{\prod_{n=1}^\infty \left(1-e^{-\pi v L n/\uptau}\right)}\sum_{n\in \mathbb{Z}}e^{-\pi v Ln^2/(2K\uptau)} \epp
\end{equation}
This result can also be expressed in terms of Dedekind eta and Jacobi theta functions. Several works have looked at the large-$L$ scaling of this partition function and have compared with Bethe ansatz results for open chains at low temperatures, see e.g.~Refs.~\cite{Sirker_openxxz} and\ \cite{KozlowskiPozsgay}. However, we are interested in performing a different limit, namely $\uptau\to \infty$, which is not apparent from the above formula. To access this limit, it is convenient to express the partition function \eqref{eq:cftZ1} in a different form. Using the Poisson re-summation formula (or, equivalently, the known modular properties of eta and theta functions), we eventually obtain
\begin{equation}\label{eq:cftZ2}
  Z_{\rm cyl}(L,\uptau) \;=\; \sqrt{K}\,\frac{e^{-E L \uptau-eL+\pi \uptau/(24 v L)}}{\prod_{n=1}^{\infty} \left(1-e^{-\pi \uptau n/(vL)}\right)} \sum_{n\in \mathbb{Z}} e^{-\pi K\uptau n^2/(2vL)} \epp
\end{equation}
Note the appearance of a prefactor $\sqrt{K}$. In the CFT literature, such factors are sometimes called $g$-factors, whose importance has been pointed out in Ref.~\cite{AffleckLudwig}. In general, they depend on the CFT as well as on the type of conformal boundary condition, which is Dirichlet here. 
Using Eq.~\eqref{eq:cftZ2}, formula \eqref{eq:overlapincft} for the overlap simplifies to
\begin{equation}\label{eq:cftresult}
  \mathcal{O}_L \;=\; \lim_{\uptau\to \infty} \frac{[Z(L,\uptau)]^2}{Z(L,2\uptau)} \;=\; \sqrt{K}e^{-e L}\epp
\end{equation}
Of course, this derivation takes only the contributions from low-energy degrees of freedom into account. In particular, the Hamiltonian \eqref{eq:xxzopenbos} is not exact any more at high energies. Treating high-energy effects would require to add multi-boson interactions and to put a cut-off on the large but finite number of bosons, which would make exact computations much harder. It is expected, but by no means obvious, that the prefactor $\sqrt{K}$ will remain unaffected by their presence, whereas the coefficient $e$ of the exponential decay will change. Lattice effects will also be responsible for additional but subleading corrections to Eq.~\eqref{eq:cftresult}. 

These observations make it desirable to explicitly check the validity of the boundary CFT formula \eqref{eq:cftresult} for the case of the spin-1/2 XXZ chain. In Ref.~\cite{Stephan2011} this was done exactly at the free fermion point $\Delta=0$, i.e.~$K=1$, where $\mathcal{O}_L=2^{-L/2}$ \cite{Mazza}, as well as numerically for other values of $\Delta$ by means of exact diagonalization techniques for small systems up to $L=28$. Agreement within a few percent was found. However, since there is now an exact Bethe ansatz-based formula for the overlap available \cite{XXZOverlap}, which is valid for any system size $L$, an analytical treatment of the large system size scaling of $\mathcal{O}_L$ is possible. In the next section, we extract the leading terms of its asymptotic expansion. This enables us to analytically confirm the presence of the universal $\sqrt{K}$ term (analytically in a certain sense, as discussed below). We also explicitly determine the coefficient of the exponential decay, which does not coincide, as expected, with the order one contribution $e$ to the ground state energy of an open XXZ chain. Furthermore, we compute the leading finite-size correction, which shows an universal algebraic decay.

\section{Asymptotics from the exact finite-size determinant formula}\label{sec:Asymptotics_from_BA} 

The starting point of our asymptotic analysis is a recently derived Bethe ansatz-based determinant formula for the overlap of the N\'eel state with an XXZ eigenstate \cite{XXZOverlap}, valid for any system size $L$. Before presenting the details of this analysis, we first outline in Sec.~\ref{sec:properties_XXZ_model} some important properties of the XXZ model and, in particular, of its ground state. In Sec.~\ref{sec:determinant_formula} we provide the explicit form of the determinant formula in terms of so-called Bethe roots. Subsequently, in Secs.~\ref{sec:asymptotics_prefactor} and \ref{sec:asymptotics_ratio_determinants}, we analyze the large system size scaling of the overlap. We calculate the thermodynamic limit (TDL) $L\to\infty$ of properly defined quantities, e.g.~the decay rate, the order one term, and (the exponent of) the leading finite-size correction, which altogether characterize the asymptotic behavior of the overlap as predicted by Eqs.~\eqref{eq:OL_result_planar}--\eqref{eq:OL_result_axial}.

\subsection{Properties of the XXZ model and its ground state}\label{sec:properties_XXZ_model}

The low-energy spectrum of the XXZ model \eqref{eq:xxzham} with anisotropy $-1 < \Delta \leq 1$ becomes gapless in the TDL, whereas it remains gapped for $\Delta>1$. Eigenstates can be explicitly determined within the framework of coordinate Bethe ansatz \cite{Bethe,GaudinCaux}. They depend on a set of parameters (Bethe roots) which satisfy, due to periodic boundary conditions, a set of so-called Bethe equations. Depending on the value of $\Delta$, this set of equations looks slightly different. We distinguish three cases: 
\begin{itemize}
	\item the `planar regime' $-1<\Delta<1$ with parameterization $\Delta=\cos\gamma$, $0<\gamma<\pi$, 
	\item the `isotropic' (also `phase-separating') point $\Delta=1$, and
	\item the `axial regime' $\Delta > 1$ with parameterization $\Delta=\cosh\eta$, $\eta>0$. 
\end{itemize}
The corresponding Bethe equations read as follows,
\begin{subequations}\label{eq:BE}
\begin{align}
&\text{planar:} & 
  \left[\frac{\sinh(\lambda_j+\frac{i\gamma}{2})}{\sinh(\lambda_j-\frac{i\gamma}{2})}\right]^L &\,=\; - \prod_{k=1}^{M}\frac{\sinh(\lambda_j-\lambda_k+i\gamma)}{\sinh(\lambda_j-\lambda_k-i\gamma)}\epc &j=1,\ldots,M\epc\label{eq:BE_planar}\\[1ex]
&\text{isotropic:} &   \left[\frac{\lambda_j+\frac{i}{2}}{\lambda_j-\frac{i}{2}}\right]^L \quad &\,=\; - \prod_{k=1}^{M}\frac{\lambda_j-\lambda_k+i}{\lambda_j-\lambda_k-i}\epc &j=1,\ldots,M\epc\label{eq:BE_isotropic}\\[1ex]
&\text{axial:} &   \left[\frac{\sin(\lambda_j+\frac{i\eta}{2})}{\sin(\lambda_j-\frac{i\eta}{2})}\right]^L &\,=\; - \prod_{k=1}^{M}\frac{\sin(\lambda_j-\lambda_k+i\eta)}{\sin(\lambda_j-\lambda_k-i\eta)}\epc &j=1,\ldots,M\epp\label{eq:BE_axial}
\end{align}
\end{subequations}
Although the Bethe equations of these three cases look similar, the structures of their solutions show perceptible differences. In the axial regime, for instance, real parts of Bethe roots are bounded in the TDL, whereas they are unbounded in the two other cases (see e.g.~Eq.~\eqref{eq:Bethe_roots_scaling} below). In the planar regime, the dependence of a Bethe state on Bethe roots shows a periodicity of $\pi$ in imaginary direction. At the isotropic point, i.e.~$\gamma\to 0$ in $\Delta=\cos\gamma$, this periodicity gets lifted after rescaling all Bethe roots by $\gamma$. 

The XXZ ground state is determined by $L/2$ many different Bethe roots $\{\lambda_j\}_{j=1}^{L/2}$ (no magnetic field) that are symmetrically distributed on the real line with distances between them as small as possible. We therefore only need to know the positive roots $\{\lambda_j\}_{j=1}^{L/4}$ where, by definition, $0 < \lambda_1 < \ldots < \lambda_{L/4}$ (remember that $L$ is assumed to be a multiple of four). For later purposes, we show in \ref{app:EMF} that the smallest and largest positive ground state Bethe roots scale in the large system size limit as 
\begin{equation}\label{eq:Bethe_roots_scaling}
  \lambda_1 \;\simeq\; 
  \left\{ \begin{array}{cl}
  \displaystyle \frac{\gamma}{L} & \text{(planar)}\\[3ex]
  \displaystyle \frac{1}{L} & \text{(isotropic)}\\[3ex]
  \displaystyle \frac{\pi}{2K\mspace{-1.5mu}(k)}\frac{\eta}{L}\enspace & \text{(axial)}\end{array}\right. 
  \epc\qquad
  \lambda_{L/4} \;\simeq\; 
  \left\{ \begin{array}{cl}
  \displaystyle \frac{\gamma}{\pi}\ln\!\Big(\frac{2L}{\pi}\Big) & \text{(planar)}\\[3ex]
  \displaystyle \frac{1}{\pi}\ln\!\Big(\frac{2L}{\pi}\Big) & \text{(isotropic)}\\[3ex]
  \displaystyle \frac{\pi}{2} \,-\, \frac{\pi}{2kK\mspace{-1.5mu}(k)}\frac{\eta}{L}\enspace & \text{(axial)}\end{array}\right.\epp
\end{equation}
In the expressions of the axial regime ($\Delta > 1$), $K(k)$ is the elliptic integral of the first kind with modulus $k$, which is uniquely determined by the anisotropy $\Delta=\cosh(\eta)$ via $\eta = \pi K(k)/K(\sqrt{1-k^2})$. Here, $K(k)$ should not be confused with the Luttinger parameter $K$ of Eq.~\eqref{eq:K}, defined for $-1 < \Delta \leq 1$ only. We explicitly see from this result that the positive ground state Bethe roots in the planar regime and at isotropic point are unbounded in the TDL, $\lambda_{L/4}\sim \ln(2L/\pi)$, while they are bounded by $\pi/2$ in the axial regime, approaching this bound in the TDL. 

The discrete distribution of ground state Bethe roots (`ground state distribution'),
\begin{equation}\label{eq:rhoL_discrete}
  \rho_L(\lambda) \;=\; \frac{1}{L}\sum_{j=1}^{L/2} \delta(\lambda-\lambda_j) \;=\; \frac{1}{L}\sum_{j=1}^{L/4} \big[\delta(\lambda-\lambda_j) + \delta(\lambda+\lambda_j)\big]\epc
\end{equation}
is a symmetric function, which is normalized to $1/2$ by definition. It becomes smooth in the TDL, 
\begin{equation}\label{eq:rho_infty}
  \rho(\lambda) \;=\; \lim_{L\to\infty} \rho_L(\lambda) \;=\;
  \left\{ \begin{array}{lll}
  \displaystyle \frac{1}{2\gamma\cosh\!\big(\frac{\pi\lambda}{\gamma}\big)}\epc & \lambda\in\mathbb{R} & \text{(planar)}\\[3ex]
  \displaystyle \frac{1}{2\cosh(\pi\lambda)}\epc & \lambda\in\mathbb{R}  & \text{(isotropic)}\\[3ex]
  \displaystyle \frac{1}{2\pi} \sum_{n\in \mathbb{Z}}\frac{\cos(2 n \lambda)}{\cosh(\eta n)}\epc\;\; & \lambda\in [-\tfrac{\pi}{2},\tfrac{\pi}{2}]\enspace  & \text{(axial)}\end{array}\right. \epp
\end{equation}
The explicit form of $\rho$ has been known for a long time (as solution of a linear integral equation (LIE) \cite{Walker59, YangYang2, Takahashi_book}), though a rigorous derivation of the LIE (at half filling) was lacking. Recently, in Ref.~\cite{Kozlowskicondensation}, a proof of this LIE was found.

\begin{figure}[tbp]
\begin{center}
  \includegraphics[width=\columnwidth]{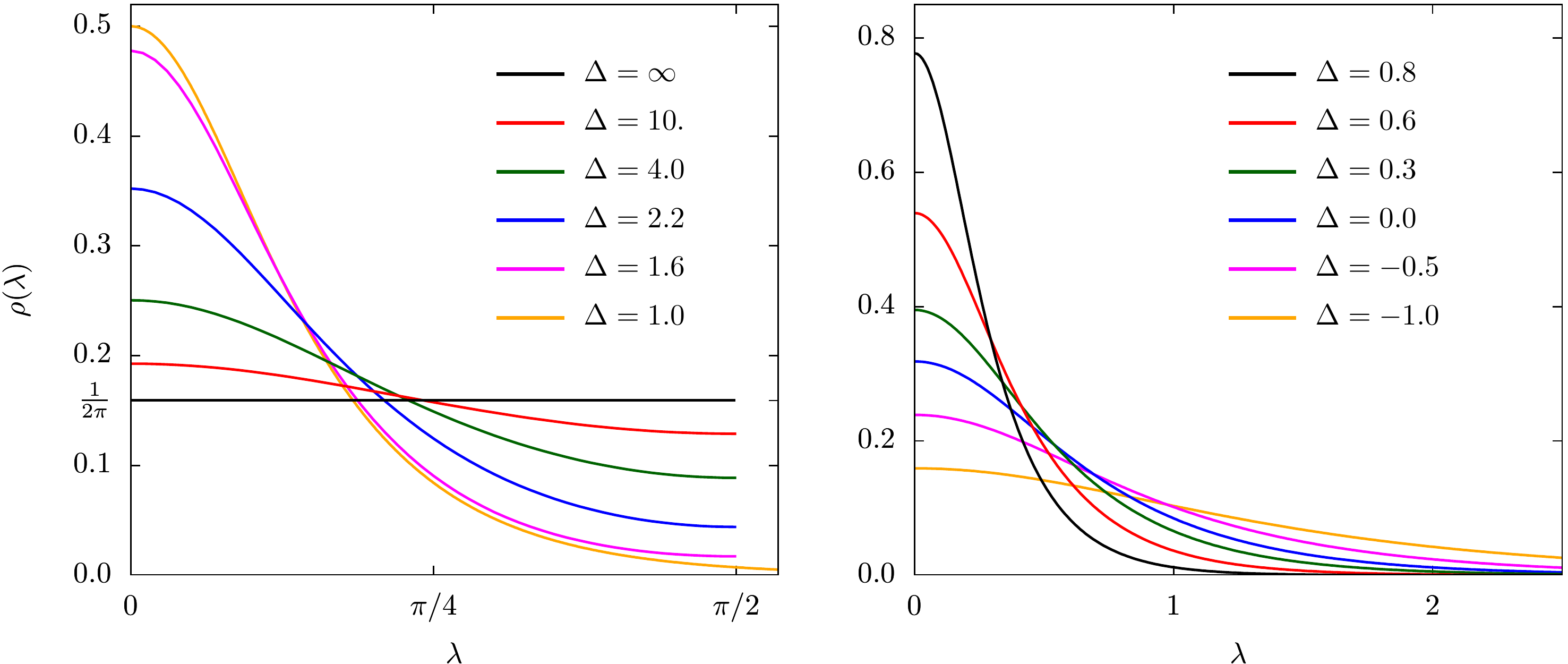}
  \caption{XXZ ground state distributions $\rho(\lambda)$ in the thermodynamic limit (TDL) for several anisotropies of the axial regime (left) and of the planar regime (right). All functions are normalized, $\int\rho(\lambda)\: {\rm d}\lambda = 1/2$. Since they are symmetric, only positive $\lambda$ are shown. The case $\Delta=1$ is compared to the distributions of the axial regime.}
  \label{fig:rhos}
\end{center}
\end{figure}

In Fig.~\ref{fig:rhos} we show some TDL ground state distributions for several values of the anisotropy parameter $\Delta$. The symmetry $\rho_L(-\lambda)=\rho_L(\lambda)$ is a necessary condition for the overlap with the N\'eel state not to vanish. If an XXZ Bethe state does not fulfil the ``condition of parity invariance'' \cite{XXZOverlap}, i.e.~$\{\lambda_j\}_{j=1}^{L/2} = \{-\lambda_j\}_{j=1}^{L/2}$, then the exact finite-size determinant formula yields exactly zero, as was shown in Ref.~\cite{XXZOverlap_Odd}. 

Despite all these differences (see e.g.~the ground state distributions \eqref{eq:rho_infty} of the planar, isotropic, and axial cases), many Bethe ansatz-based formulas look similar for the three cases (see e.g.~Eqs.~\eqref{eq:BE} and Ref.~\cite{QISMbook}). They can be transformed into each other by replacing all hyperbolic sine functions (planar regime) by sine functions (axial regime) or by its argument (isotropic point), and so forth. This also holds for the exact finite-size overlap formula whose explicit expression is presented in the next subsection.

\subsection{Finite-size determinant formula}\label{sec:determinant_formula}

We exploit the determinant formula of Ref.~\cite{XXZOverlap} for the overlap of an XXZ Bethe state with the N\'eel state by inserting the positive ground state Bethe roots $\{\lambda_j\}_{j=1}^{L/4}$. In all three cases (planar, isotropic, axial), the exact overlap formula has the following structure (`prefactor' $\mathcal{P}_L$ times `ratio of determinants' $\mathcal{R}_L$),
\begin{subequations}\label{eq:determinant_formula}
\begin{equation}\label{eq:determinant_formula_structure}
 \mathcal{O}_L \;=\; \mathcal{P}_L\,\cdot\,\mathcal{R}_L \qquad\text{with}\qquad \mathcal{P}_L \;=\; \prod_{j=1}^{L/4} p(\lambda_j)\,,\quad \mathcal{R}_L \;=\; \left|\frac{\det_{L/4}(G^{+})}{\det_{L/4}(G^{-})}\right|\epc
\end{equation}
where the `prefactor function' $p$ is positive for real arguments and reads
\begin{equation}\label{eq:prefactor_function}
p(\lambda)\;=\; 
  \left\{ \begin{array}{ll}
  \displaystyle \frac{\tanh(\lambda+\frac{i\gamma}{2}) \tanh(\lambda-\frac{i\gamma}{2})}{4\sinh^2(2\lambda)}\enspace & \text{(planar)}\\[3ex]
  \displaystyle \frac{\lambda^2+\frac{1}{4}}{16\lambda^2}  & \text{(isotropic)}\\[3ex]
  \displaystyle \frac{\tan(\lambda+\frac{i\eta}{2}) \tan(\lambda-\frac{i\eta}{2})}{4\sin^2(2\lambda)} & \text{(axial)}\end{array}\right. \epp
\end{equation}
The matrices $G^{\pm}$ are of dimension $L/4$ and given by
\begin{align}\label{eq:G_matrices}
G_{jk}^\pm &\;=\; \delta_{jk} \,+\, \frac{K_1^\pm(\lambda_j,\lambda_k)}{L \tilde{\rho}_L(\lambda_j)}\epc \quad j,k=1,\ldots,L/4 \epc\\[1.5ex]
 \tilde{\rho}_L(\lambda) &\;=\; K_{1/2}(\lambda) \,-\, \frac{1}{L}\sum_{l=1}^{L/4}K_1^+(\lambda,\lambda_l)\epc\label{eq:rho_L_tilde}
\end{align}
with `kernel functions' $K_1^\pm(\lambda,\mu)=K_1(\lambda-\mu) \pm K_1(\lambda+\mu)$ and 
\begin{equation}\label{eq:kernels}
K_\beta(\lambda)\;=\; 
  \left\{ \begin{array}{ll}
  \displaystyle \frac{1}{2\pi}\,\frac{\sin(2\beta\gamma)}{\sinh^2(\lambda)+\sin^2(\beta\gamma)}\enspace & \text{(planar)}\\[3ex]
  \displaystyle \frac{1}{2\pi}\,\frac{2\beta}{\lambda^2+\beta^2}  & \text{(isotropic)}\\[3ex]
  \displaystyle \frac{1}{2\pi}\,\frac{\sinh(2\beta\eta)}{\sin^2(\lambda)+\sinh^2(\beta\eta)} & \text{(axial)}\end{array}\right. \epc\qquad \beta \;=\; 1/2,\, 1\epp
\end{equation}
\end{subequations}
It can be shown \cite{Takahashi_book,Kozlowskicondensation} that $\lim\nolimits_{L\to\infty}\tilde{\rho}_L(\lambda) = \rho(\lambda)$ with the TDL ground state distribution $\rho$ of Eq.~\eqref{eq:rho_infty}. 

We are interested in the large system size scaling of this overlap $\mathcal{O}_L$, which we expect to be, in the three cases `planar', `isotropic', and `axial', of the form \eqref{eq:OL_result_planar}, \eqref{eq:OL_result_isotropic}, and \eqref{eq:OL_result_axial}, respectively. To show this, we first of all observe that the prefactor $\mathcal{P}_L$ alone accounts for the exponential decay $e^{-\alpha L}$ of $\mathcal{O}_L$. This will become manifest in our analysis in Sec.~\ref{sec:asymptotics_ratio_determinants}, where we show that $\mathcal{R}_L$ converges to a constant and, hence, contributes at most to the order one term. Thus, the decay rate can be expressed as $\alpha = -\lim_{L\to\infty}\ln(\mathcal{P}_L)/L$. In \ref{app:decay_rate} we explicitly compute it for all $\Delta \geq -1$.

For later convenience, we consider the logarithm of the overlap $\mathcal{O}_L$ plus $L$ times the decay rate $\alpha$, 
\begin{equation}\label{eq:exact_overlap_formula}
 \ln \mathcal{O}_L \,+\, \alpha L \;=\; \ln \mathcal{P}_L \,+\, \alpha L \,+\, \ln \mathcal{R}_L  \;=\; \sum_{j=1}^{L/4} \ln[p(\lambda_j)] \,+\, \alpha L \,+\, \ln\!\left|\frac{\det_{L/4}(G^{+})}{\det_{L/4}(G^{-})}\right|\epp
\end{equation} 
In the following, we calculate the order one contributions as well as the leading finite-size corrections of $\ln(\mathcal{P}_L)+\alpha L$ and $\ln(\mathcal{R}_L)$ explicitly, proving that Eqs.~\eqref{eq:OL_result_planar}--\eqref{eq:OL_result_axial} are indeed the correct asymptotic expansions of $\mathcal{O}_L$ for large system size $L$. In Sec.~\ref{sec:asymptotics_prefactor} we study the different contributions of the `prefactor term' $\ln(\mathcal{P}_L) + \alpha L$ while Sec.~\ref{sec:asymptotics_ratio_determinants} is devoted to the analysis of the $\ln(\mathcal{R}_L)$-term. In Sec.~\ref{sec:results_overlap_BA} we summarize the results by comparing the contributions of the prefactor term with those of the ratio of determinants. We further identify the source of the leading finite-size correction of the overlap in the three different cases `planar', `isotropic', and `axial'.

\subsection{Asymptotic analysis of the prefactor}\label{sec:asymptotics_prefactor}

Let us define the `prefactor term' $P(L)$ by
\begin{equation}\label{eq:P(L)}
  P(L) \;=\; \ln(\mathcal{P}_L) \,+\, \alpha L \;=\; \sum_{j=1}^{L/4} \ln[p(\lambda_j)] \,+\, \alpha L\epc
\end{equation}
where $p$ is the prefactor function of Eq.~\eqref{eq:prefactor_function}, and $\alpha$ is the decay rate as computed in \ref{app:decay_rate}. We are interested in an asymptotic large-$L$ expansion of $P(L)$, 
\begin{equation}\label{eq:P(L)_asymp}
  P(L) \;\simeq\; P_0 \,+\, P_1(L) \,+\,\ldots\epc
\end{equation}
where we refer to the zeroth order term $P_0$ as the order one contribution of $P(L)$ and to the first order term $P_1$ as its leading finite-size correction. The dots denote higher orders, i.e.~subleading finite-size corrections. 

In Sec.~\ref{sec:Ps_order_one_contribution} we derive an explicit expression for $P_0$ in all three cases (planar, isotropic, axial), which is the most difficult part and which will be the main result of our asymptotic analysis. Furthermore, in Sec.~\ref{sec:corrections} we extract the leading finite-size correction $P_1(L)$, which turns out to be qualitatively different in the three cases `planar', `isotropic', and `axial'. In the planar regime, for instance, we will see that the prefactor term scales as $P(L) \simeq \ln(K) + A/L^\delta$ with an algebraically decaying finite-size correction $P_1(L) = A/L^\delta$ whereas, at the isotropic point ($K=1/2$), it scales as $P(L) \simeq -\ln(2) + \mathcal{O}(\ln^{-2}(L))$. In the axial regime, on the contrary, the correction to the zeroth order term $P_0=-2\ln(2)$ is exponentially small, $P(L) \simeq -2\ln(2) + \mathcal{O}(e^{-cL})$.

\subsubsection{Order one term.}\label{sec:Ps_order_one_contribution}

One idea to compute the order one contribution $P_0$ of the prefactor term \eqref{eq:P(L)_asymp} could be to use the Euler-Maclaurin formula. By means of this formula, a sum over ground state Bethe roots as in Eq.~\eqref{eq:P(L)} can be approximated for large system size $L$ by an integral (times $L$) plus order one and subleading corrections (see e.g.~Ref.~\cite{Woynarovich89} and Eq.~\eqref{eq:EMF} in \ref{app:EMF}). In our case, the function under the integral, $\ln[p(\lambda)]$, shows logarithmic singularities at the boundaries of the integration interval. In \ref{app:EMF} we demonstrate that an application of the Euler-Maclaurin formula to such a function encounters technical difficulties. Nevertheless, by making use of the expansion of the log-gamma function \cite{Pearce07,Costin08}, an analytic treatment of the singularity at the lower boundary is possible in all three cases (planar, isotropic, axial). In contrast, the singularity at the upper boundary can be treated only for the axial case. For the planar and isotropic cases, unfortunately, the singularity of $\ln[p(\lambda)]$ at infinity leads to insurmountable difficulties. This is why this approach works well in the axial regime, while it does not work in the planar regime and at the isotropic point. 

Another approach to compute sums over Bethe roots is based on the well-established technique of non-linear integral equations (NLIE). This technique was intended for and successfully applied to the computation of finite-temperature thermo\-dynamics \cite{Kluemper92,Kluemper93}. It was shown that it is also applicable to finite systems (of arbitrary length) at zero temperature \cite{Kluemper91}. More specifically, the NLIE can be formulated for any excited (low-lying) state of the finite system. ``Zero temperature'' then just means to consider the ground state. This approach works well in all three cases (planar, isotropic, axial). The only difference is that the driving terms and kernels of the NLIE look slightly different \cite{Bortz05}. The structure of the NLIE, however, as well as the logic of the derivation of the order one term $P_0$ are exactly the same. Note that our analysis of the NLIE does not only provide an exact expression for the order one contribution $P_0$, but also enables us to determine the scaling of the leading finite-size correction $P_1(L)$ (see Sec.~\ref{sec:corrections}). 

In the following, we focus on the second approach in the regime $0\leq \Delta < 1$. The other part of the planar regime, $-1<\Delta<0$, as well as the isotropic point $\Delta=1$ and the axial regime $\Delta>1$ can be treated similarly. The NLIE approach is based on the following formulas \cite{Damerau07,Kluemper91}. We consider the `auxiliary function'
\begin{equation}\label{eq:def_a}
  \mathfrak{a}(\lambda) \;=\; \left[\frac{\sinh(\lambda-\frac{i\gamma}{2})}{\sinh(\lambda+\frac{i\gamma}{2})}\right]^L \prod_{k=1}^{M}\frac{\sinh(\lambda-\lambda_k+i\gamma)}{\sinh(\lambda-\lambda_k-i\gamma)}\epp
\end{equation}
Obviously, due to the Bethe equations \eqref{eq:BE_planar}, $\mathfrak{a}(\lambda_j)=-1$. Furthermore, we consider a contour $\mathcal{C}$ in the complex plane that encircles the Bethe roots $\{\lambda_j\}_{j=1}^{M}$ in a counter-clockwise manner and that does not encircle hole solutions of the Bethe equations (if present, e.g.~for excited states of the XXZ chain). However, since we are interested in the ground state of the XXZ model, characterized by roots only, we do not need to be concerned too much about hole solutions. For a function $f$ that is analytic inside the contour $\mathcal{C}$ the sum of $f(\lambda_j)$ over all Bethe roots $\lambda_j$ can be then expressed by the following contour integral, 
\begin{equation}\label{eq:sum_over_BR}
  \sum_{j=1}^{M} f(\lambda_j) \;=\; \oint_{\mathcal{C}} \frac{{\rm d} \omega}{2\pi i}\:\frac{\mathfrak{a}'(\omega)}{1+\mathfrak{a}(\omega)}\,f(\omega)\epp
\end{equation}
The auxiliary function $\mathfrak{a}$ satisfies the following non-linear integral equation (NLIE) \cite{Kluemper91,Damerau07}, 
\begin{equation}\label{eq:NLIE_a}
  \ln[\mathfrak{a}(\lambda)] \;=\; i\gamma L \,+\, L\ln\!\left[\frac{\sinh(\lambda-\frac{i \gamma}{2})}{\sinh(\lambda+\frac{i \gamma}{2})}\right] \,+\, i\oint_{\mathcal{C}} K_1(\lambda-\omega)\ln\!\left[1+\mathfrak{a}(\omega)\right]{\rm d}\omega \epc
\end{equation}
where $K_1$ is the kernel function of Eq.~\eqref{eq:kernels}. In principle, the full $L$-dependence of a sum over Bethe roots is completely determined by the last two equations. In practice, however, it is a difficult task (if not impossible) to solve the NLIE analytically, in order to get an explicit expression for $\mathfrak{a}(\lambda)$, even in the limit of large system size $L\to\infty$. 

In the case of the ground state, which consists of $L/2$ symmetrically distributed real Bethe roots, it is convenient to choose the contour $\mathcal{C}$ such that the entire real axis is enclosed. This way, $\mathcal{C}$ becomes independent of the detailed positions of the ground state Bethe roots and, in particular, independent of~$L$. The $L$-dependence of the sum in Eq.~\eqref{eq:sum_over_BR} is fully captured by the `driving term' of the NLIE \eqref{eq:NLIE_a}. Still, an analytical solution of the NLIE is out of reach. Due to the complex contour $\mathcal{C}$, even a numerical evaluation is problematic. This second obstacle can be cleared by expanding the contour $\mathcal{C}$ to an infinitly long rectangle with edges parallel to the real axis (at levels $\pm i\gamma/2$) and by defining two new auxiliary functions,
\begin{equation}\label{eq:def_b_bbar}
  \mathfrak{b}(x) \;=\; \mathfrak{a}^{-1}(x-i\gamma/2)\epc\qquad \bar{\mathfrak{b}}(x) \;=\; \mathfrak{a}(x+i\gamma/2)\epc\qquad x\in\mathbb{R}\epp 
\end{equation}
Using this specific contour and plugging these definitions into Eq.~\eqref{eq:NLIE_a}, a set of two coupled non-linear integral equations (NLIEs) for the functions $\mathfrak{b}$ and $\bar{\mathfrak{b}}$ can be derived \cite{Kluemper91}. Since the contributions from the vertical parts of the rectangular contour can be neglected, it reads
\begin{subequations}\label{eq:NLIEs_planar}
\begin{align}\label{eq:NLIEs_planar_1}
  \ln[\mathfrak{b}(x)] &\;=\; L\cdot d(x) \,+\, \big[F\ast \ln(1+\mathfrak{b})\big](x) \,-\, \big[F^-\ast \ln(1+\bar{\mathfrak{b}})\big](x)\epc\\[1.2ex]
  \ln[\bar{\mathfrak{b}}(x)] &\;=\; L\cdot d(x) \,+\, \big[F\ast \ln(1+\bar{\mathfrak{b}})\big](x) \,-\, \big[F^+\ast \ln(1+\mathfrak{b})\big](x)\epc
\end{align}
where $(f\ast g)(x) = \int_{-\infty}^\infty f(x-y)g(y)\,{\rm d}y$ denotes a convolution, and where the driving term and the integration kernels are given by 
\begin{align}
  d(x) &\;=\; \ln\!\left[\tanh\!\left(\frac{\pi x}{2\gamma}\right)\right]\epc \label{eq:driving_term}\\[0.8ex]
  F(x) &\;=\; \int_{-\infty}^\infty \frac{\sinh\!\left((\frac{\pi}{2}-\gamma)k\right)e^{ikx}}{2\cosh\!\left(\frac{k\gamma}{2}\right)\sinh\!\left((\pi-\gamma)\frac{k}{2}\right)}\;\frac{{\rm d}k}{2\pi} \epc\qquad F^\pm(x)=F(x\pm i\gamma \mp i0^+) \label{eq:F-kernel}\epp
\end{align}
\end{subequations}
Now, we would like to rewrite the right-hand side of Eq.~\eqref{eq:sum_over_BR} in terms of the auxiliary functions $\mathfrak{b}$ and $\bar{\mathfrak{b}}$. First of all, we note that the driving term $d$ of these new NLIEs is related to the TDL ground state distribution $\rho$ via $id'(x+i\gamma/2)=2\pi\rho(x)$. 
Provided that $f$ is analytic in the whole strip $\{\lambda\in\mathbb{C} \;| \,-\frac{\gamma}{2} < {\rm Im}(\lambda) < \frac{\gamma}{2}\, \}$ the sum over ground state Bethe roots can be explicitly written as 
\begin{multline}\label{eq:sum_over_BR_b}
  \sum_{j=1}^{L/2} f(\lambda_j) \;=\; L\mspace{-2mu}\int_{-\infty}^\infty \rho(x)f(x)\:{\rm d}x \,+ \int_{-\infty}^\infty \frac{{\rm d}x}{2\pi i}\: \Big\{\big[(\delta-F)\ast\ln'(1+\mathfrak{b})\big](x)\, f(x-\tfrac{i\gamma}{2})\\  \,-\, \big[(\delta-F)\ast\ln'(1+\bar{\mathfrak{b}})\big](x)\, f(x+\tfrac{i\gamma}{2})\Big\}\epp
\end{multline} 
Here, $\delta$ is the delta distribution defined by $(\delta\ast g)(x) = g(x)$. The derivation of Eq.~\eqref{eq:sum_over_BR_b} is straightforward. First, we deform the contour $\mathcal{C}$ in Eq.~\eqref{eq:sum_over_BR} to the rectangle described above Eqs.~\eqref{eq:def_b_bbar} and subsequently insert the definitions \eqref{eq:def_b_bbar} themselves. Since on the lower contour (lower edge of the rectangle) the inverse of $\mathfrak{a}$ is used in the definition of $\mathfrak{b}$, also the derivative of $\ln(\mathfrak{b})$ occurs. In order to express everything by derivatives of $\ln(1+\mathfrak{b})$ and $\ln(1+\bar{\mathfrak{b}})$ solely, we plug in the derivative of the NLIE \eqref{eq:NLIEs_planar_1} with respect to $x$. After an expedient shift of the integration contour in one of the four convolution terms, we finally arrive at Eq.~\eqref{eq:sum_over_BR_b}. In the very last step we explicitly used the analyticity condition of $f$ mentioned above Eq.~\eqref{eq:sum_over_BR_b}.

We further note that ${\rm Re}(\mathfrak{b}) = {\rm Re}(\bar{\mathfrak{b}})$ and ${\rm Im}(\mathfrak{b}) = -{\rm Im}(\bar{\mathfrak{b}})$, which reflects the symmetry of the ground state Bethe roots and basically means that the set \eqref{eq:NLIEs_planar} of two NLIEs can be reduced to one NLIE for one unknown function $\mathfrak{b}$. However, solving the whole set \eqref{eq:NLIEs_planar} is computationally not much more costly, especially if we take into account that the real parts of $\mathfrak{b}$ and $\bar{\mathfrak{b}}$ are symmetric functions and that the imaginary parts are antisymmetric. If we require that $f(x\pm i\gamma/2)$ as a function of real argument $x$ has the same symmetry as the auxiliary functions $\mathfrak{b}$ and $\bar{\mathfrak{b}}$, i.e.~${\rm Re}\,f(x\pm i\gamma/2) = {\rm Re}\,f(-x\pm i\gamma/2)$ and ${\rm Im}\,f(x\pm i\gamma/2) = -{\rm Im}\,f(-x\pm i\gamma/2)$, the sum over positive ground state Bethe roots can be simplified to 
\begin{multline}\label{eq:sum_over_BR_sym}
  \sum_{j=1}^{L/4} f(\lambda_j)  \;=\; L\mspace{-2mu}\int_{0}^\infty \!\rho(x)f(x)\:{\rm d}x \,+\, {\rm Im}\mspace{-2mu}\int_{-\infty}^\infty \big[(\delta-F)\ast\ln'(1+\mathfrak{b})\big](x) f(x-\tfrac{i\gamma}{2})\,\frac{{\rm d}x}{2\pi}\epp
\end{multline} 
Since we sum over positive ground state Bethe roots only, we integrate from zero to infinity in the first term, and an additional factor of $1/2$ appears in the second term. 

Regarding an evaluation of Eqs.~\eqref{eq:NLIEs_planar} and \eqref{eq:sum_over_BR_b} or \eqref{eq:sum_over_BR_sym}, there are three big advantages as compared to Eqs.~\eqref{eq:sum_over_BR} and \eqref{eq:NLIE_a}. Firstly, both functions $\mathfrak{b}$ and $\bar{\mathfrak{b}}$ have to be evaluated only for purely	real arguments. Secondly, they are very small close to the origin ($\sim x^L$), as can be most easily seen from their definitions, i.e.~Eqs.~\eqref{eq:def_b_bbar} together with Eq.~\eqref{eq:def_a}. Thirdly, the integrations are convolutions along the real axis. To evaluate them we can apply efficient numerical techniques such as the Fast Fourier Transform algorithm. We solve the NLIEs \eqref{eq:NLIEs_planar} for a fixed system size $L$ by iteration, which converges after a few steps (a couple of hundreds or less) and provides the functions $\mathfrak{b}$ and $\bar{\mathfrak{b}}$ to high numerical accuracy. Since $L$ enters only through the prefactor of the driving term $d$, we are able to compute solutions for system sizes up to $L=10^{10}$ or even larger. Subsequently, after an integration by parts, we plug the solution $\mathfrak{b}$ into Eq.~\eqref{eq:sum_over_BR_sym} and compute the integral. In this sense, we can calculate sums over ground state Bethe roots exactly, i.e.~to high numerical accuracy (up to $10^{-14}$ if we want), for any system size $L$ and, in particular, for very large $L$. This enables us, for instance, to calculate the order one term $P_0$ of the prefactor term $P(L)$ as well as its leading finite size correction $P_1(L)$, both to high numerical accuracy (for details see the following discussion as well as Sec.~\ref{sec:corrections} and, in particular, Fig.~\ref{fig:delta_of_Delta} therein).

We are interested in the special case
\begin{equation}\label{eq:f=ln(p)}
  f(x) \;=\; \ln[p(x)] \qquad\text{with}\quad\; p(x) \;=\; \frac{\sinh^2(x) + \sin^2(\frac{\gamma}{2})}{4\big[\!\sinh^2(x) + \cos^2(\frac{\gamma}{2})\big]\sinh^2(2x)}\epp
\end{equation}
We observe that the function $f$ satisfies the symmetry requirements (see statement right before Eq.~\eqref{eq:sum_over_BR_sym}), but it does not satisfy the analyticity condition (see right before Eq.~\eqref{eq:sum_over_BR_b}). The function $f'$ has a pole on the real axis at zero (with residue $-2$), and therefore $f$ is not analytical at zero and shows a branch cut similar to that of the function $\ln(x)$. This leads to an additional explicit term which can be easily computed by considering the contour integration in Eq.~\eqref{eq:sum_over_BR} directly. We integrate the right-hand side by parts and expand the contour $\mathcal{C}$ to the rectangle that eventually leads to the $\mathfrak{b}$-$\bar{\mathfrak{b}}$-formulation \eqref{eq:NLIEs_planar} of the NLIE. This order of steps is most convenient. Though, the other way around leads, of course, to the same result. Then, from this pole at zero, we encounter an explicit term which has to be subtracted from the right-hand side of Eq.~\eqref{eq:sum_over_BR_sym}. Therefore, Eq.~\eqref{eq:sum_over_BR_sym} will be modified by the following extra term, 
\begin{equation}\label{eq:extra_term_ln(2)}
 \frac{1}{2} \oint_{D_\epsilon(0)} \frac{{\rm d}\omega}{2\pi i}\;f'(\omega)\ln[1+\mathfrak{a}(\omega)] \;=\; -\ln[1+\mathfrak{a}(0)] \;=\; -\ln(2)\epc
\end{equation}
where $D_\epsilon(0)$ is the disc around zero with small radius $\epsilon$, and where we used that $\mathfrak{a}(0)=1$ which is due to the symmetry of ground state Bethe roots, $\{\lambda_j\}_{j=1}^{L/2}=\{-\lambda_j\}_{j=1}^{L/2}$. Taking this term into account and using that the first integral in Eq.~\eqref{eq:sum_over_BR_sym} is the decay rate $\alpha$ (see \ref{app:decay_rate}), we eventually obtain
\begin{equation}\label{eq:P(L)_result}
  P(L) \;=\; -\ln(2) \,+\, {\rm Im}\mspace{-2mu}\int_{-\infty}^\infty \big[(\delta-F)\ast\ln'(1+\mathfrak{b})\big](x)\: \ln\!\left[p(x-\tfrac{i\gamma}{2})\right]\frac{{\rm d}x}{2\pi} \epp
\end{equation}
This formula gives us access to the prefactor term $P(L)$ for any system size $L$, either numerically to very high accuracy (as discussed before) or, in a certain sense, analytically in the TDL (as discussed in the following).

The NLIEs \eqref{eq:NLIEs_planar} can be analyzed in the TDL, similarly to the low-temperature analysis of Ref.~\cite{Kluemper98}. One observation from our numerical investigation of the NLIEs \eqref{eq:NLIEs_planar} is that the functions $\ln'(1+\mathfrak{b})$ and $\ln'(1+\bar{\mathfrak{b}})$ change significantly only in a small region around $\pm \mathcal{L}$ with $\mathcal{L} = \frac{\gamma}{\pi}\ln(2L)$. Furthermore, they vanish for $x\to 0$ (rapidly) and for $x\to\infty$ (slowly), i.e.~$\ln'(1+\mathfrak{b})$ and $\ln'(1+\bar{\mathfrak{b}})$ show `bumps' around $\pm\mathcal{L}$. Thus, it is convenient to define `scaling functions' by
\begin{equation}\label{eq:b_bbar_scaled}
  \mathfrak{b}_{\pm}(x) \;=\; \mathfrak{b}(\pm(x+\mathcal{L}))\epc \qquad \bar{\mathfrak{b}}_{\pm}(x) \;=\; \bar{\mathfrak{b}}(\pm(x+\mathcal{L}))\epc\qquad \mathcal{L} \;=\; \frac{\gamma}{\pi}\ln(2L)\epp
\end{equation}
Using these definitions in the NLIEs \eqref{eq:NLIEs_planar} and neglecting contributions from convolution integrals which connect bumps of $\ln'(1+\mathfrak{b})$ or $\ln'(1+\bar{\mathfrak{b}})$ around $\mathcal{L}$ with those around $-\mathcal{L}$, we obtain the following NLIEs for the scaling functions,
\begin{subequations}\label{eq:NLIEs_scaling}
\begin{align}
  \ln[\mathfrak{b}_\pm(x)] &\;=\; -e^{-\pi x/\gamma} \,+\, \big[F\ast \ln(1+\mathfrak{b}_\pm)\big](x) \,-\, \big[F^-\ast \ln(1+\bar{\mathfrak{b}}_\pm)\big](x)\epc\\[0.8ex]
  \ln[\bar{\mathfrak{b}}_\pm(x)] &\;=\; -e^{-\pi x/\gamma} \,+\, \big[F\ast \ln(1+\bar{\mathfrak{b}}_\pm)\big](x) \,-\, \big[F^+\ast \ln(1+\mathfrak{b}_\pm)\big](x)\epc
\end{align}
\end{subequations}
where the integration kernels are the same as in Eq.~\eqref{eq:F-kernel}. Note that the driving terms of these NLIEs are independent of $L$. We interpret the scaling functions (i.e.~solutions of these `TDL-NLIEs') as the TDL of the auxiliary functions $\mathfrak{b}$ and $\bar{\mathfrak{b}}$. Taking into account that the function $\ln(p)$ scales for large arguments as 
\begin{equation}\label{eq:scaling_ln(p)}
  \ln\!\left[p(x-\tfrac{i\gamma}{2})\right] \;\stackrel{x\,\to\,\pm\infty}{\longrightarrow}\; -\ln\!\left[\sinh^2(2x)\right] \,\pm\, 2i\gamma\epc
\end{equation}
we derive from Eq.~\eqref{eq:P(L)_result} the following expression, 
\begin{equation}\label{eq:P_0}
  P_0 \;=\; \lim_{L\to\infty} P(L) \;=\; -\ln(2) \,+\, \frac{2\gamma K\ln(2)}{\pi} \,+\, \frac{4K}{\pi}\int_{-\infty}^{\infty} {\rm Im}\big\{ \ln[1+\mathfrak{b}_+(y)]\,\big\}\: {\rm d}y\epp
\end{equation}
Here, we used $\int_{-\infty}^\infty [\delta(x)-F(x)] {\rm d}x = K$ with the Luttinger parameter $K=\pi/[2(\pi-\gamma)]$ and, for the second term in Eq.~\eqref{eq:P_0}, that $\lim_{x\to\pm\infty}\ln[1+\mathfrak{b}(x)] = \ln(2)$. To derive the third term on the right-hand side of Eq.~\eqref{eq:P_0}, we used an integration by parts, the definition \eqref{eq:b_bbar_scaled}, and that $\lim_{x\to\pm\infty} \ln'[\sinh^2(2x)] = 4$. The result \eqref{eq:P_0}, with $\mathfrak{b}_+$ as the (unique) solution of Eqs.~\eqref{eq:NLIEs_scaling}, is exact. Un\-for\-tun\-ate\-ly, we were unable to simplify it further. However, by solving the TDL-NLIEs \eqref{eq:NLIEs_scaling} nu\-mer\-i\-cal\-ly to high accuracy (essentially machine precision), we could show that the integral in Eq.~\eqref{eq:P_0} is given by
\begin{equation}\label{eq:nice_Imb+_formula}
  \int_{-\infty}^{\infty} {\rm Im}\big\{ \ln[1+\mathfrak{b}_+(y)]\,\big\}\: {\rm d}y \;=\; -\frac{\gamma\ln(2)}{2} \,-\, \frac{\pi-\gamma}{2}\ln\!\left(\frac{\pi-\gamma}{\pi}\right)\epp
\end{equation}
This eventually leads to the final result of this subsection, 
\begin{equation}\label{eq:P_0_result_planar}
  P_0 \;=\; \ln(K) \qquad \text{(planar, isotropic)}\epp
\end{equation}
The limit $\gamma\to 0$ (isotropic point, $K=1/2$) can be easily taken and yields $P_0 = -\ln(2)$. 

In the axial regime, the analysis of the prefactor term $P(L)$ is similar but less involved. This is due to the fact that Bethe roots are bounded and integrations (of the con\-vo\-lu\-tions and of all other integrals) run from $-\pi/2$ to $\pi/2$. Hence, the integral in Eq.~\eqref{eq:P(L)_result} is subleading, rather than of order one as in the planar regime, and does not contribute to $P_0$. Instead, there is a second additional explicit term, which is similar to that of Eq.~\eqref{eq:extra_term_ln(2)} and stems from the pole of $\ln'(p)$ at $\pi/2$ (also with residue $-2$), 
\begin{equation}\label{eq:extra_term_ln(2)_b}
  \frac{1}{2}\oint_{D_\epsilon\left(\!\frac{\pi}{2}\!\right)} \frac{{\rm d}\omega}{2\pi i}\;f'(\omega)\ln[1+\mathfrak{a}(\omega)] \;=\; -\ln[1+\mathfrak{a}(\pi/2)] \;=\; -\ln(2)\epp
\end{equation}
Here, we used that in the axial regime the hyperbolic sine functions in the definition \eqref{eq:def_a} of the auxiliary function $\mathfrak{a}$ are replaced by usual sine functions. Hence, $\mathfrak{a}(\pi/2)=1$, again due to the symmetry of ground state Bethe roots $\{\lambda_j\}_{j=1}^{L/2}=\{-\lambda_j\}_{j=1}^{L/2}$. Therefore, we obtain for the order one term $P_0$ of the prefactor term $P(L)$ in the axial regime the following explicit, rigorously derived expression, 
\begin{equation}\label{eq:P_0_result_axial}
  P_0 \;=\; -2\ln(2) \qquad \text{(axial)}\epc
\end{equation}
which is independent of the anisotropy parameter $\Delta>1$.

\subsubsection{Leading finite-size corrections.}\label{sec:corrections}

The analysis of the leading finite-size correction $P_1(L)$ of the prefactor term $P(L)$ proceeds as follows. There are a priori two main contributions to $P_1(L)$, which we discuss separately. The first comes from the terms that connect bumps of the auxiliary functions around $\mathcal{L}$ with bumps around $-\mathcal{L}$, which were neglected in the derivation of the TDL-NLIEs \eqref{eq:NLIEs_scaling}. Shifting  the integration variables of the convolution integrals by $\pm \mathcal{L}$ and inserting the definition \eqref{eq:b_bbar_scaled} of the scaling functions into the NLIEs \eqref{eq:NLIEs_planar}, one observes that the leading correction to the TDL-NLIEs \eqref{eq:NLIEs_scaling} stems from convolution terms with $\ln(1+\mathfrak{b}_-)$, $\ln(1+\bar{\mathfrak{b}}_-)$ (not from those with $\ln(1+\mathfrak{b}_+)$, $\ln(1+\bar{\mathfrak{b}}_+)$), see e.g.~Eqs.~(6)--(8) of Ref.~\cite{Kluemper98}. It is therefore of the form
\begin{equation}
  \int_{-\mathcal{L}}^\infty F(x+y+2\mathcal{L})\ln[1+\mathfrak{b}_+(y)]\:{\rm d}y \;\sim\; e^{-2\mathcal{L}\beta}\epc 
\end{equation}
where $\beta$ is the exponent of the leading exponential decay of the kernel function $F$. The proportionality factor is a function of $x$ and, to leading order, independent of $\mathcal{L}$. So, it is unimportant to us. In summary, the leading corrections to the scaled auxiliary functions are of order $\exp\{-2\mathcal{L}\beta\}$. The second contribution is an integral proportional to $F(\mathcal{L}) \simeq \exp\{-\mathcal{L}\beta\}$, which was neglected in going from  Eq.~\eqref{eq:P(L)_result} to its scaled version \eqref{eq:P_0}. Hence, the leading finite-size corrections of the scaled auxiliary functions are subleading with respect to this integral and, therefore, $P_1(L)$ is proportional to $\exp\{-\mathcal{L}\beta\}$ with $\mathcal{L} = \gamma\ln(2L)/\pi$. The exponent $\beta$ can be easily obtained by investigating the poles of the Fourier transform of $F$ in the complex plane, see Eq.~\eqref{eq:kernels}. The result is simply $\beta= \text{min}\{\frac{\pi}{\gamma},\frac{2\pi}{\pi-\gamma}\}$.

Thus, by inspecting the decay of the kernel function $F$ of the NLIEs \eqref{eq:NLIEs_planar}  for large arguments, we have shown that the leading finite-size correction $P_1(L)$ of the prefactor term $P(L)$ decays algebraically as $A/L^\delta$ with amplitude $A=A(\Delta)$ and exponent\footnote{Note that \textit{a priori} this amplitude and this exponent can be different from the amplitude and the exponent of the leading finite-size correction of $\ln(\mathcal{O}_L)$ (in the planar regime), cf.~Eq.~\eqref{eq:OL_result_planar}. As we will discuss later in Sec.~\ref{sec:results_overlap_BA}, they are in fact the same. We therefore denote them by the same letters here. }
\begin{equation}\label{eq:delta}
  \delta \;=\; \delta(\Delta) \;=\; \text{min}\left\{1,\frac{2\gamma}{\pi-\gamma}\right\}\epp
\end{equation}
This formula is one of the main results of this subsection. It is valid in the entire planar regime $-1<\Delta=\cos(\gamma)<1$. 

\begin{figure}
\begin{center}
  \includegraphics[width=\columnwidth]{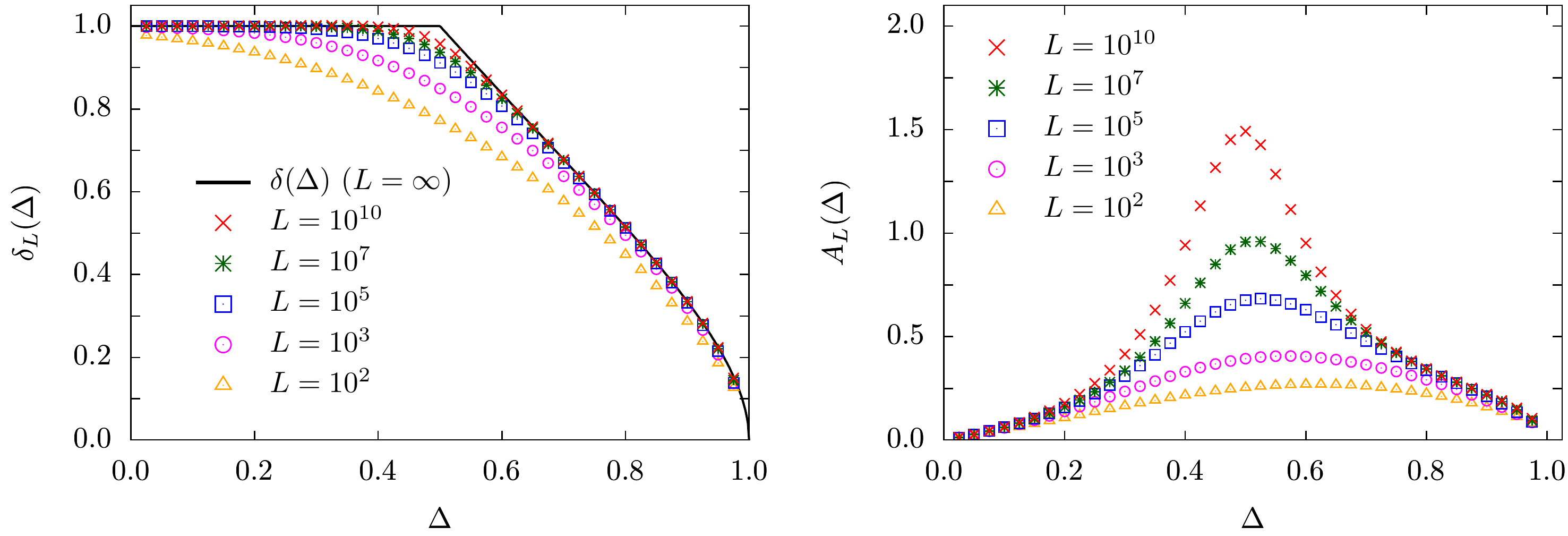}
  \caption{Exponent $\delta$ and amplitude $A$ of the leading finite-size cor\-rec\-tion of the prefactor term $P(L)$ in the planar regime as functions of $\Delta$. The prediction \eqref{eq:delta} for $\delta(\Delta)$ (see black line in the left panel) is compared to data of the finite-size exponent $\delta_L$ (Eq.~\eqref{eq:delta_L}, dots). Computations of $\delta_L$ as well as of finite-size amplitudes $A_L$ (Eq.~\eqref{eq:A_L}, dots in the right panel) are based on solutions of NLIE (see text).}
  \label{fig:delta_of_Delta}
\end{center} 
\end{figure}
In Fig.~\ref{fig:delta_of_Delta} we show the exponent $\delta=\delta(\Delta)$ as a function of the anisotropy $\Delta$ in the regime $0<\Delta<1$ (black solid line). It is compared to `finite-size exponents' 
\begin{equation}\label{eq:delta_L}
  \delta_L \;=\; -\frac{\ln[P(L+\Delta L) - P_0] - \ln[P(L) - P_0]}{\ln(L+\Delta L) - \ln(L)}\epc
\end{equation}
which we computed by means of the NLIEs \eqref{eq:NLIEs_planar} for several values of $L$ up to $L=10^{10}$ (with a properly chosen $\Delta L \ll L$). We further checked for several values of $\Delta$ with $-1<\Delta<0$ (by solving Bethe equations) that the exponents $\delta_L$ are very close to one already for $L \gtrsim 100$ (results not shown in the figure). We also computed `finite-size amplitudes' (see right panel of Fig.~\ref{fig:delta_of_Delta}),
\begin{equation}\label{eq:A_L}
  A_L \;=\; \exp\left\{\frac{\ln(L+\Delta L)\ln[P(L) - P_0] - \ln(L)\ln[P(L+\Delta L) - P_0]}{\ln(L+\Delta L) - \ln(L)}\right\}\epc
\end{equation}
again to very high accuracy by solving the NLIEs \eqref{eq:NLIEs_planar}. 

Furthermore, we can take the isotropic limit of the kernel function $F$. For large arguments $x$ it decays algebraically, $F(x) \sim 1/x^2$, instead of exponentially as in the planar regime. Following the line of arguments of the planar case, we infer that the leading finite-size correction $P_1(L)$ at the isotropic point is logarithmic, $P_1(L) \simeq \mathcal{O}(\ln^{-2}(L))$. We verified this by solving the corresponding non-linear integral equations up to $L=10^{15}$. Note that solving the Bethe equations \eqref{eq:BE_isotropic} up to $L \approx 8000$ is by far not sufficient to reach the asymptotic regime (see lower curve of the right panel of Fig.~\ref{fig:corrections_planar_isotropic} in Sec.~\ref{sec:asymptotics_ratio_determinants} and, in particular, the inset). 

In the axial regime, on the other hand, the leading finite-size correction $P_1(L)$ is exponentially small, $P_1(L) \sim e^{-cL}$. Thus, it is sufficient to solve the Bethe equations \eqref{eq:BE_axial} for system sizes up to $L=320$ and to estimate the value of $c$ by an extrapolation (see e.g.~Sec.~\ref{sec:asymptotics_ratio_determinants} and, in particular, the inset of Fig.~\ref{fig:logO}).

\subsection{Asymptotic analysis of the ratio of determinants}\label{sec:asymptotics_ratio_determinants} 

This section deals with the asymptotics of the ratio of determinants $\mathcal{R}_L$, Eq.~\eqref{eq:determinant_formula_structure}, for large system size $L$. 
The formula for the ratio of determinants reads (see Sec.~\ref{sec:determinant_formula})
\begin{equation}\label{eq:ratiodet1}
  \mathcal{R}_L=\frac{\det_{1\leq j,k\leq L/4}\left(L \tilde{\rho}_L(\lambda_j)\delta_{jk}+K_1^+(\lambda_j,\lambda_k)\right)}{\det_{1\leq j,k\leq L/4}\left(L \tilde{\rho}_L(\lambda_j)\delta_{jk}+K_1^-(\lambda_j,\lambda_k)\right)}\epc
\end{equation}
where the kernel functions $K_1^\pm$ and the function $\tilde{\rho}_L$ are given by Eqs.~\eqref{eq:kernels} and \eqref{eq:rho_L_tilde}, respectively. It is valid for all system sizes $L$ (with the only restriction that $L$ has to be a multiple of four). We first demonstrate how $\mathcal{R}_L$ converges to a ratio of Fredholm determinants in the TDL, which we evaluate analytically using a theorem of Ref.~\cite{Basor97}. We also would like to draw the interested reader's attention to Ref.~\cite{BottcherSilbermann} for a review on the asymptotic analysis of such types of determinants. Subsequently, the finite-size corrections to this TDL value are studied numerically by solving the Bethe ansatz equations \eqref{eq:BE} for finite but large system sizes up to $L=8192$.

Let us first focus on the planar and isotropic cases, which are the most complicated. In the TDL, the distribution of ground state Bethe roots becomes a smooth function $\rho$ on the real line, Eq.~\eqref{eq:rho_infty}, which is the limit of the sequence of functions $\tilde{\rho}_L$ \cite{Kozlowskicondensation}. Neglecting the finite-size corrections, which we shall analyze at the end of this subsection, Eq.~\eqref{eq:ratiodet1} may be recast as a limit of ratio of two Fredholm determinants (denoted in the following by `Det' to distinguish it from the finite-size determinant, which is denoted by `det'),
\begin{subequations}
\begin{equation}\label{eq:fredholm1}
  \lim_{L\to \infty}\mathcal{R}_L\;=\;\lim_{\Lambda\to \infty} \,\frac{\Det(1 + T_\Lambda[\hat{K}_1] + H_\Lambda[\hat{K}_1])}{\Det(1 + T_\Lambda[\hat{K}_1] - H_\Lambda[\hat{K}_1])}\,\epp
\end{equation}
The integral operators $T_\Lambda[\hat{K}_1]$ and $H_\Lambda[\hat{K}_1]$ act on functions in $L^2([0,\Lambda])$ as follows,
\begin{align}
  (T_\Lambda[\hat{K}_1] f)(\lambda) &\;=\; \int_0^\Lambda K_1(\lambda-\mu)f(\mu)\;{\rm d}\mu\epc \label{eq:Toeplitzact}\\[1.2ex]
  (H_\Lambda[\hat{K}_1] f)(\lambda) &\;=\; \int_0^\Lambda K_1(\lambda+\mu)f(\mu)\;{\rm d}\mu\epp \label{eq:Hankelact}
\end{align}
\end{subequations}
Here, $K_1(\lambda)$ is the kernel defined in Eq.~\eqref{eq:kernels}. The discussion below applies to any real-valued kernel under reasonable smoothness assumptions. In analogy to regular matrices, we used the notation $T$ for Toeplitz (resp. $H$ for Hankel) for the part that depends only on the difference (resp. sum) of the arguments $\lambda$ and $\mu$. The symbol $\hat{K}_1$ stands for the cosine transform of $K_1$, $\hat{K}_1(q)=2\int_0^\infty K_1(\lambda)\cos(q\lambda)\:{\rm d}\lambda$. Said differently, $T_\Lambda[\hat{K}_1]$ and $H_\Lambda[\hat{K}_1]$ act as shown in Eqs.~\eqref{eq:Toeplitzact} and \eqref{eq:Hankelact}, where $K_1$ is the inverse cosine transform of $\hat{K}_1$. Such types of determinants have been widely studied in the mathematical literature. In particular, the limit \eqref{eq:fredholm1} follows from Theorem 5 of Ref.~\cite{Basor97}. We obtain
\begin{equation}\label{eq:limRL}
  \lim_{L\to\infty} \ln(\mathcal{R}_L ) \;=\; 2\lim_{\Lambda\to\infty}{\rm Tr}\,\left\{ H_\Lambda\big[\ln(1+\hat{K}_1)\big] \right\} \;=\; \frac{1}{2}\ln\!\left[1+\hat{K}_1(0)\right]\epp
\end{equation}
Computing the integral $\hat{K}_1(0)=2\int_0^\infty K_1(\lambda)\,{\rm d}\lambda$ in Eq.~\eqref{eq:limRL} is straightforward. We find
\begin{equation}
  \lim_{L\to \infty}\ln(\mathcal{R}_L) \;=\; \frac{1}{2}\ln\!\left[2\left(1-\frac{\gamma}{\pi}\right)\right] \;=\; -\frac{1}{2}\ln(K)\epc
\end{equation}
where $K$ is the Luttinger parameter of Eq.~\eqref{eq:K}. We note in passing that the large $\Lambda$ limit of each determinant in (\ref{eq:fredholm1}) may also be determined separately using the aforementioned theorem. We find (for any smooth integral operator $V$)
\begin{equation*}
  \Det(1+T_\Lambda[\hat V]\pm H_\Lambda[\hat V])\underset{\Lambda\to \infty}{\sim}\exp\left(\Lambda S(0)\pm\frac{1}{4}\hat{S}(0)+\frac{1}{2}\int_{0}^{\infty} \lambda S(\lambda)^2\,{\rm d}\lambda\right)\epc
\end{equation*}
with $\hat{S}(q)=\ln[1+\hat V(q)]$ and the inverse cosine transform $S(\lambda)=\frac{1}{\pi}\int_0^\infty \hat{S}(q)\cos (q\lambda)\,{\rm d}q$.

In the axial case, the algebra is only slightly different. Formula \eqref{eq:fredholm1} still holds but the kernel now acts on $L^2([0,\pi/2])$ instead of $L^2([0,\Lambda])$. Another difference is that both determinants are finite. We obtain
\begin{equation}
 \lim_{L\to \infty}\ln(\mathcal{R}_L) \;=\; \ln\!\left[1+\frac{1}{\pi}\int_0^{\pi/2}\!K_1(\lambda)\:{\rm d}\lambda\right] \;=\; \ln(2)\epp
\end{equation}
Note that even though the kernel itself depends on $\Delta=\cosh(\eta)$, see Eq.~\eqref{eq:kernels}, the final result in the axial regime does not. To summarize, we have found
\begin{equation}\label{eq:lnR_asymp}
  \lim_{L\to\infty}\mathcal{R}_L \;=\; \left\{\begin{array}{ll} 1/\sqrt{K} \quad & \text{(planar and isotropic)}\\[1ex] \phantom{-}2 &\text{(axial)}   \end{array}\right.\epp
\end{equation}
In all three cases, the order one contribution is half of the order one contribution of $\ln(\mathcal{P}_L)$ with opposite sign (see Sec.~\ref{sec:Ps_order_one_contribution}).

We also performed numerical checks of these formulas by computing the exact ratio \eqref{eq:fredholm1} by means of the Gaussian quadrature method, demonstrated in Ref.~\cite{Bornemann} and successfully applied, for instance,  in Ref.~\cite{Dugave2014} in the context of low-temperature, large-distance asymptotics of two-point functions of the XXZ model. In the planar regime, the kernels are regular and decay very fast at infinity. We find that the ratio of Fredholm determinants can be evaluated to almost arbitrary accuracy. For example, choosing $\Lambda=30$ and a few hundred points for the Gaussian quadrature, typically allows to confirm Eq.~\eqref{eq:lnR_asymp} to more than $30$ digits of numerical accuracy. The situation becomes less favorable when approaching the isotropic limit from below. In fact, exactly at the isotropic point the kernel decays algebraically, and we were able to confirm the result $\lim_{L\to \infty} \mathcal{R}_L=\sqrt{2}$ only up to relative error of the order $10^{-7}$. Finally, in the axial case, the numerical evaluation of the ratio of Fredholm determinants becomes tremendously accurate, presumably due to the fact that the kernels now act on a finite interval. This holds even close to the isotropic point. Choosing, for instance, $\Delta=1.01$ and $300$ quadrature points yields about $50$ digits of accuracy. 

\begin{figure}
\begin{center}
  \includegraphics[width=\columnwidth]{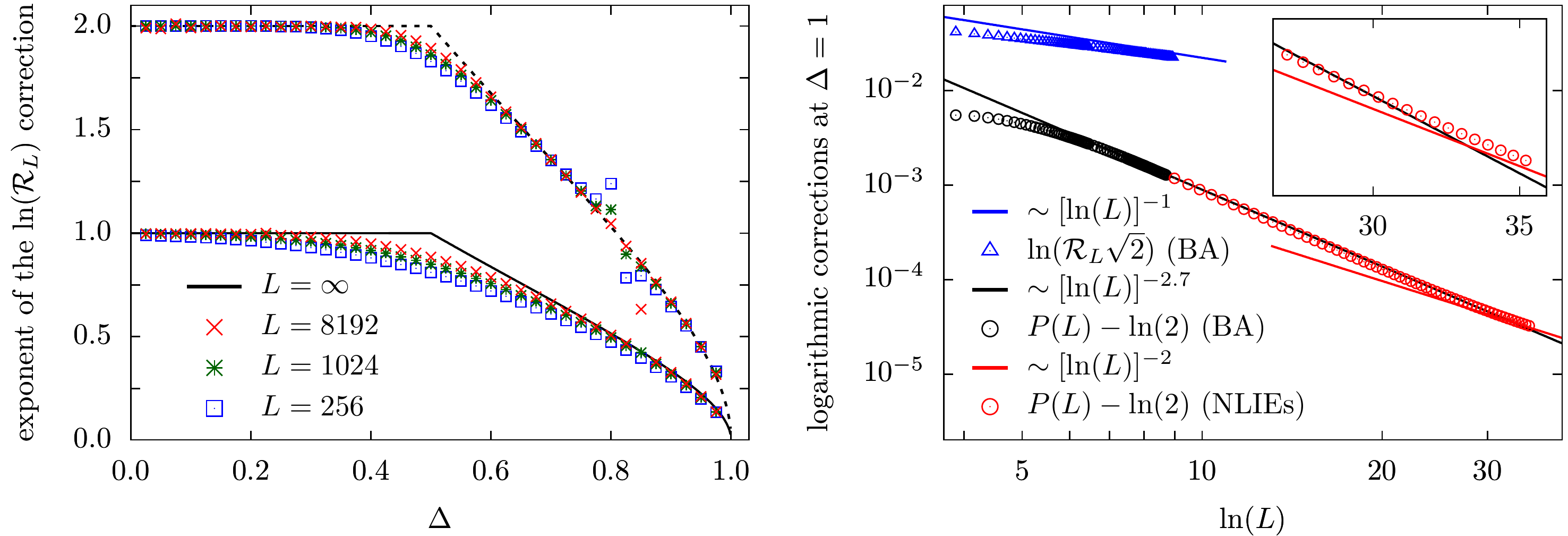}
  \caption{Left: Exponent of the leading finite-size correction of $\ln(\mathcal{R}_L)$. The conjecture $2\delta$ (black dotted line) is compared to finite-size exponents (upper curves), obtained by solving Bethe equations for finite system size $L$ and inserting the solution into $\ln(\mathcal{R}_L)+\ln(\sqrt{K})$. For a comparison, the finite-size exponents $\delta_L$ of the prefactor term $P(L)$ are shown as well (lower curves, see Fig.~\ref{fig:delta_of_Delta}). Right: Leading finite-size corrections at the isotropic point $\Delta=1$ as functions of $\ln(L)$ in a double-logarithmic plot. Data for $P(L)+\ln(2)$ (lower curve) as well as for $\ln(\mathcal{R}_L)-\ln(\sqrt{2})$ (upper curve) are compared to straight lines, whose slopes indicate the exponents of the algebraic decay, $\sim [\ln(L)]^{-\alpha}$, $\alpha = 1,2$, of the finite-size correction of $\ln(\mathcal{R}_L)$, $P(L)$, respectively. Note that the asymptotics of the latter only sets in for very large $L \gtrsim 10^{12}$ (see inset).}
  \label{fig:corrections_planar_isotropic}
\end{center} 
\end{figure}

Another way to approach the ratio of determinants $\mathcal{R}_L$ is to compute it directly by solving the Bethe equations \eqref{eq:BE} and inserting the solution of the ground state Bethe roots into the exact finite-size formula \eqref{eq:ratiodet1}. This procedure is limited by the computational time needed to solve the Bethe equations to a given accuracy. We solved Bethe equations of the three different cases up to $L=8192$. This way, we could numerically confirm that $\mathcal{R}_L$ scales indeed to the TDL result \eqref{eq:lnR_asymp}. Furthermore, at the same time, we got access to the finite-size corrections of $\ln(\mathcal{R}_L)$. In the planar regime, we could numerically verify  (see left panel of Fig.~\ref{fig:corrections_planar_isotropic}; regime $-1<\Delta<0$ not shown in the figure) that the finite-size correction behaves as $\sim L^{-2\delta}$, with $\delta = \text{min}\{1,4K-2\}$ being the exponent of the leading finite-size correction of the prefactor term $P(L)$. 

At the isotropic point $\Delta=1$, the correction appears to be logarithmic, namely of the form $\sim 1/\ln(L)$ (see upper curve in the right panel of Fig.~\ref{fig:corrections_planar_isotropic}). In the axial regime, the finite size correction of $\ln(\mathcal{R}_L)$ is exponentially small ($\sim e^{-cL}$, see inset of Fig.~\ref{fig:logO}).

\subsection{Results}\label{sec:results_overlap_BA} 

We summarize the results of the previous subsections. In the planar regime $-1 < \Delta < 1$, we found
\begin{align}
  \ln(\mathcal{P}_L) &\;\simeq\; -\alpha L + \ln(K) + AL^{-\delta}\epc \notag\\[0.12ex]
  \ln(\mathcal{R}_L) &\;\simeq\; -\ln(\sqrt{K}) + \mathcal{O}(L^{-2\delta})\epc \notag\\[0.12ex]
  \Rightarrow\quad \ln(\mathcal{O}_L) &\;\simeq\; -\alpha L + \ln(\sqrt{K}) + A L^{-\delta}\epp \label{eq:overlap_result_planar}
\end{align}
Here, the decay rate $\alpha$ is determined by Eq.~\eqref{eq:alpha_planar} and $K$ is the Luttinger parameter of Eq.~\eqref{eq:K}. The exponent $\delta$, Eq.~\eqref{eq:delta}, as well as the amplitude $A$ describe the algebraic decay of the leading finite-size correction. Fig.~\ref{fig:delta_of_Delta} illustrates the behavior of the two parameters $\delta$ and $A$. The leading finite-size correction of $\ln(\mathcal{R}_L)$ is completely dominated by the leading correction of $\ln(\mathcal{P}_L)$ (see left panel of Fig.~\ref{fig:corrections_planar_isotropic}). 
By exponentiating the asymptotic expression for $\ln(\mathcal{O}_L)$, we finally obtain the result \eqref{eq:OL_result_planar}. 

The asymptotic formula \eqref{eq:overlap_result_planar} still holds in the isotropic limit, with the only difference that the leading finite-size correction is no longer algebraic but logarithmic. Since the correction of the prefactor term is of order $\mathcal{O}(\ln^{-2}(L))$, the leading correction of the overlap $\ln(\mathcal{O}_L)$ stems here from the ratio of determinants, which is\ \;$\sim 1/\ln(L)$ (see left panel of Fig.~\ref{fig:corrections_planar_isotropic}). 

\begin{figure}
\begin{center}
  \includegraphics[width=\columnwidth]{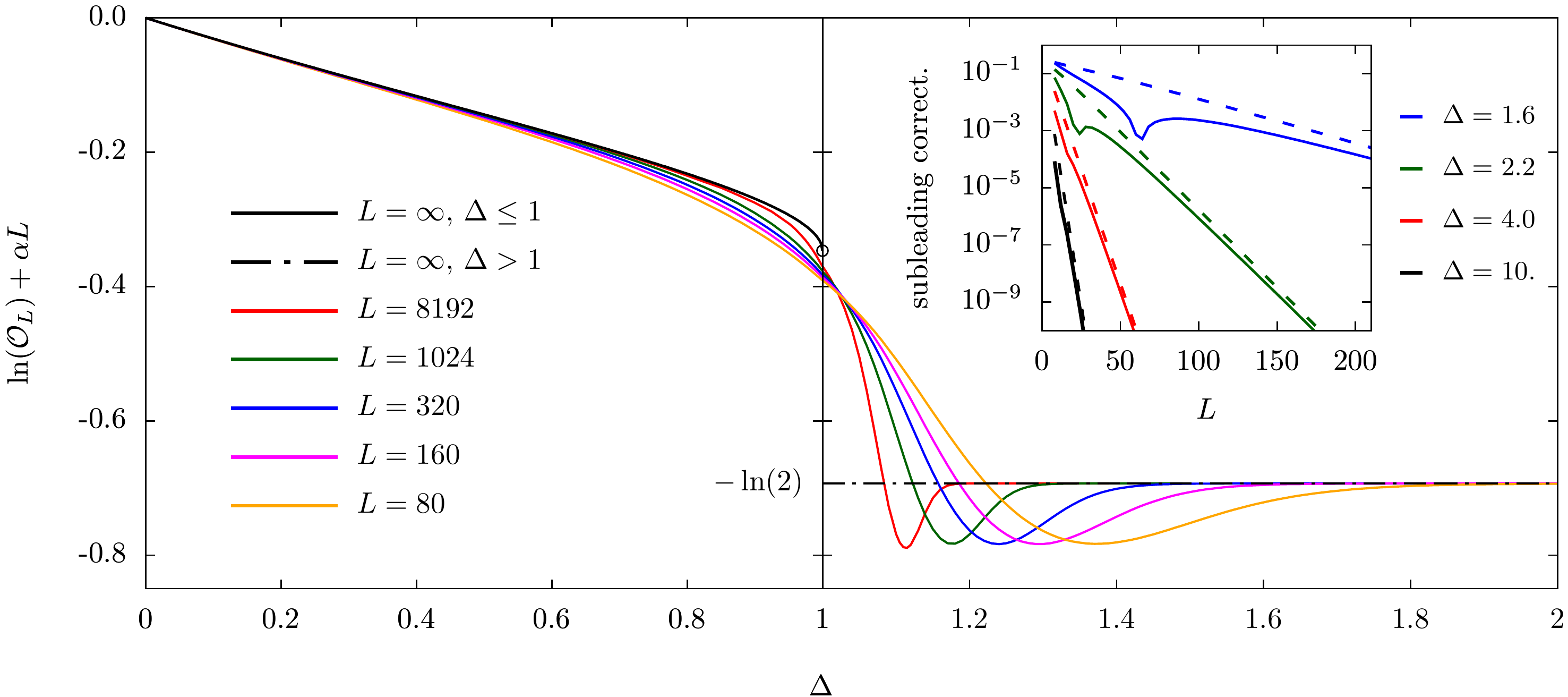}
  \caption{Logarithm of the overlap $\mathcal{O}_L$ plus $L$ times $\alpha$ as a function of the anisotropy $0\leq\Delta\leq 2$ for different system sizes $L$. The solid black curve depicts the order one term in the planar regime, $\ln(\sqrt{K})$, which is the TDL of $\ln(\mathcal{O}_L)+\alpha L$. The circle at the phase-separating point $\Delta=1$ represents the value $-\ln(\sqrt{2})$. The horizontal dotted black line at level\ \,$-\ln(2)$ shows the $\Delta$-independent order one term in the axial (gapped) regime. Inset: exponentially small corrections in the axial regime to the order one term (solid lines) and to the ratio of determinants (dashed lines), both with the same $c$ in $e^{-cL}$ (see text).}
  \label{fig:logO}
\end{center}
\end{figure}

The result in the axial regime $\Delta > 1$ (gapped phase) reads
\begin{align}
  \ln(\mathcal{P}_L) &\;\simeq\; -\alpha L - 2\ln(2) + \mathcal{O}(e^{-cL})\epc \notag\\[0.36ex]
  \ln(\mathcal{R}_L) &\;\simeq\; \ln(2) + \mathcal{O}(e^{-cL})\epc \notag\\[0.36ex]
  \Rightarrow\quad \ln(\mathcal{O}_L) &\;\simeq\; -\alpha L - \ln(2) + \mathcal{O}(e^{-cL})\epc \label{eq:overlap_result_axial}
\end{align}
where $\alpha$ is determined by Eq.~\eqref{eq:alpha_axial}. The leading finite-size corretion is exponentially small, with the same $c=c(\Delta)$ for the corrections of $\ln(\mathcal{P}_L)$ and $\ln(\mathcal{R}_L)$. We further observe that $\lim_{\Delta\to 1} c(\Delta)=0$ and $\lim_{\Delta\to\infty}c(\Delta) = \infty$ (see inset of Fig.~\ref{fig:logO}). The former is expected since the correction becomes logarithmic in the isotropic limit. The latter is also expected since there is no finite-size correction to the order one term in the Ising limit (see below Eq.~\eqref{eq:OL_result_axial}). Again, exponentiating the asymptotic expression for $\ln(\mathcal{O}_L)$ yields the expected result, Eq.~\eqref{eq:OL_result_axial}. 

In Fig.~\ref{fig:logO} we show the order one term (black solid and dashed lines) in all three cases (planar, isotropic, axial) together with some finite-size calculations based on solutions of Bethe equations up to $L=8192$. Note that the order one term as function of anisotropy is not continuous at $\Delta=1$, a clear signature of the phase transition.

\section{Conclusion}\label{sec:discussion} 

In this paper, we presented an extensive asymptotic analysis of the overlap between the N\'eel state and the ground state of the XXZ spin chain. We found that such an overlap always decays exponentially, which is expected for a non-trivial ground state of a quantum many body system. Nevertheless, we computed the corresponding decay rate exactly (see \ref{app:decay_rate}). Subleading corrections to this behavior are most interesting, as they reveal important physical properties of the underlying quantum system at low temperature. In the gapped case (axial regime $\Delta>1$), for instance, the order one correction is exactly $1/2$ and independent of the anisotropy $\Delta$, reflecting the symmetry broken nature of the ground state.

We put particular emphasis on the gapless phase of the model (planar and isotropic cases), where our results are perhaps most significant. In Sec.~\ref{sec:overlap_CFT} we argued (following Refs.~\cite{Stephan2009,Stephan2011}) that the order one correction is universal and exactly given by $\sqrt{K}$ with the Luttinger parameter $K$. The derivation relies on a standard, but crucial CFT assumption. In an imaginary time picture, the N\'eel state plays the role of a boundary condition which renormalizes to a conformal invariant boundary condition \cite{Cardy_bccoperatorcontent}, which is Dirichlet in our case. We checked this assumption in Sec.~\ref{sec:Asymptotics_from_BA} by means of a microscopic approach, taking full advantage of the integrability of the XXZ spin chain. This approach is technically considerably more involved, but nevertheless doable using a recently derived exact finite-size formula for the overlap \cite{XXZOverlap}. As we have shown, the order one term in the asymptotic expansion of the overlap is in perfect  agreement with the prediction of CFT. 

The evidence gathered in Sec.~\ref{sec:Asymptotics_from_BA} for the correctness of the CFT formula in the XXZ spin chain is overwhelming. 
We checked the validity of identity \eqref{eq:OL_result_planar} to machine precision (about $16$ digits) for several values of the anisotropy $\Delta$ by solving the NLIE equations \eqref{eq:NLIEs_planar} iteratively and subsequently computing the integral \eqref{eq:P(L)_result}. In this sense, the order one term is exactly given by the TDL formulas \eqref{eq:NLIEs_scaling} and \eqref{eq:P_0}, but its explicit expression \eqref{eq:P_0_result_planar} could not be rigorously derived yet (in a mathematical strict sence). Aside from this issue, the derivation of the CFT term presented here was fully analytical. Furthermore, we derived the asymptotic formula \eqref{eq:OL_result_axial} of the axial (gapped) regime, in Sec.~\ref{sec:Asymptotics_from_BA} by means of the NLIE approach as well as in \ref{app:EMF} by means of the Euler-Maclaurin formula (EMF). Again, from a mathematically rigorous perspective, there is a bottleneck to prove formula \eqref{eq:OL_result_axial} by means of the EMF. It lies in the TDL ground state root distribution \eqref{eq:rho_infty} whose applicability to sums of functions that we consider in this paper, see Eqs.~\eqref{eq:exact_overlap_formula} and \eqref{eq:prefactor_function}, is not proven yet. We refer to Ref.~\cite{Kozlowskicondensation} for a proof in case of slightly more regular functions. We hope  that further elaboration on the ideas of Ref.~\cite{Kozlowskicondensation} will close this gap (in the line of arguments of the EMF approach).

Another important feature of the order one term of the overlap is that it is universal, in the sense that a \emph{different} microscopic model that renormalizes to the \emph{same} CFT would have the \emph{same} order one term. This can be beautifully illustrated by considering the simpler Haldane-Shastry chain\cite{Haldane,Shastry} which scales to a Luttinger liquid with Luttinger parameter $K=1/2$ (the corresponding point in XXZ is the isotropic point). The overlap between the ground state and any product state may be expressed in square root determinant form \cite{MehtaMehta74,StephanPollmann}, leading to the exact N\'eel overlap 
\begin{equation}
  \mathcal{O}_L \;=\; \frac{(L/2)^{L/2}(L/2)!}{L!} \;=\; 2^{-1/2}e^{-\alpha L}\left(1+\mathcal{O}(L^{-1})\right)  
\end{equation}
with decay rate $\alpha=\ln(2)-1/2 \simeq 0.19314718$. As expected, the universal order one term is $\sqrt{K}=\sqrt{1/2}$, but the non-universal decay rate is different from the XXZ decay rate \eqref{eq:alpha_isotropic}, in agreement with the discussion above.

We further extracted the form of the leading finite-size correction of the overlap, which takes the form of a power law with an exponent that we determined exactly to be $\delta={\rm min}\{1,4K-2\}$. At the isotropic point ($\gamma=0$, $K=1/2$), this turns into an $1/\ln L$ correction. The exponent and other subleading corrections can be derived, in principle, also from a perturbed CFT analysis. To do so, one needs to identify the least irrelevant bulk/boundary operators authorized by the symmetries, and then to compute the generated corrections. Such corrections depend on the scaling dimensions of the operators considered. The leading exponent, for instance, can be explained by considering perturbations by the stress-tensor (which gives $L^{-1}$) combined with a cosine term (which gives $L^{-4K+2}$). Such calculations typically require a little bit of hindsight from the underlying microscopic model. This analysis was not necessary here. The determination of the corrections follows from a standard reasoning based on the structure of the kernels that appear in the non-linear integral equations. This approach can also be generalized to analyze higher subleading terms. We note in passing that the knowledge of these corrections might prove useful in the study of certain basis-dependent R\'enyi-Shannon entropies \cite{Stephan2009,Stephan2011,Luitz}, of which our overlap is a particular case. These entropies have also been studied using CFT, but the lack of tractable finite-size formulas makes the determination of subleading corrections even more crucial.  

There are several other potential directions in which the present work may be generalized. For simplicity, we focused our attention on the ground state solely. However, the determinant formula we relied on holds for any eigenstate, and an asymptotic analysis for low-lying energy states should be also possible. In this case, however, boundary CFT does not predict any extra contributions. Still, this could be verified using the techniques presented in this paper. Another possible generalization concerns the asymptotic analysis of the partition function of the six-vertex model on a torus or cylinder, the latter being related \cite{Tsuchiya} to the overlap we have studied. It would be interesting to try to push our analysis one step further and to derive the full CFT partition function \eqref{eq:cftZ2} as an order one correction to the partition function of the discrete model in the TDL. A similar CFT analysis may be applied also to open chains \cite{ZBM,Stephan2011}, where boundary contributions for low-lying excitations display a richer structure \cite{Bondesan_rectangle1,Bondesan_rectangle2}. This includes order one as well as power-law corrections originating from the Cardy-Peschel formula \cite{CardyPeschel}. In that case, tractable formulas for the overlap with the N\'eel state are unfortunately not available yet. Another intriguing idea would be to go back to the ground state of the periodic XXZ chain and to realize a different conformal invariant boundary condition, e.g.~a Neumann boundary condition. This can be done by computing the overlap of the ground state with a normalized equal-weight superposition of all product states \cite{Geometricentanglement}. We hope that the present paper stimulates future works in such directions.

\section*{Acknowledgements}

We are grateful to F.~G{\"o}hmann, Y.~Ikhlef, M.~Karbach, A.~Kl{\"u}mper, K.~K.~Kozlowski, P.~Pearce, and M.~Rajabpour for useful discussions. MB acknowledges financial support by the DFG research group FOR-2316 funded by the DFG.

\appendix

\section{Computation of the decay rate}\label{app:decay_rate}

The decay rate $\alpha=\alpha(\Delta)$ can be obtained by taking the TDL of the logarithm of the overlap $\mathcal{O}_L$ (or equivalently of the prefactor $\mathcal{P}_L$) divided by the system size $L$, 
\begin{equation}\label{eq:alpha_via_logP}
  \alpha(\Delta) \;=\; -\lim_{L\to\infty} \frac{\ln(\mathcal{P}_L)}{L} \;=\; -\lim_{L\to\infty} \frac{1}{L}\sum_{j=1}^{L/4}\ln[p(\lambda_j)] \;=\; -\int \ln[p(\lambda)]\rho(\lambda)\: {\rm d}\lambda\epp
\end{equation}
Hence, for any value of $\Delta \geq -1$, it is determined by the integral of the logarithm of the prefactor function $p$, Eq.~\eqref{eq:prefactor_function}, times the TDL ground state distribution $\rho(\lambda)$, Eq.~\eqref{eq:rho_infty}. This relation has been often used in the past, though a rigorous proof is still lacking for our purpose. Note, however, that such type of results can be rigorously proved \cite{Kozlowskicondensation} for arbitrary anisotropy $\Delta>-1$ and for a wide class of functions $f$ slightly more regular than the one in Eq.~\eqref{eq:alpha_via_logP}.

Let us first consider the planar regime. The positive ground state Bethe roots are unbounded in the TDL (see e.g.~Eq.~\eqref{eq:Bethe_roots_scaling} or Fig.~\ref{fig:rhos}). Therefore, we have to integrate in Eq.~\eqref{eq:alpha_via_logP} over the whole positive real axis. Thus, the decay rate reads ($\Delta = \cos\gamma$)
\begin{subequations}\label{eq:alpha_planar}
\begin{align}\label{eq:alpha_planar_a}
  \alpha(\Delta) &\;=\; -\int_0^\infty \ln\left[\frac{\sinh^2(\lambda) + \sin^2(\frac{\gamma}{2})}{4(\sinh^2(\lambda) + \cos^2(\frac{\gamma}{2}))\sinh^2(2\lambda)}\right]\,\frac{{\rm d}\lambda}{2\gamma\cosh(\frac{\pi\lambda}{\gamma})}\\[1.6ex]
  &\;=\; \ln(2) \,- \int_{-\infty}^\infty \frac{\sinh((\frac{\pi}{\gamma}-1)k)\sinh(k)}{2\sinh(\frac{\pi k}{\gamma})\cosh(2k)} \frac{{\rm d}k}{k} \epp\label{eq:alpha_planar_b}
\end{align}
\end{subequations}
In particular, for $\Delta=\Delta_m =\cos\gamma_m$ with $\gamma_m = \pi/(2m+1)$, $m\in \mathbb{N}_0$, the integral in Eq.~\eqref{eq:alpha_planar_b} can be further simplified. By doing some manipulations (i.e.~adding a factor $e^{ikx}$ under the intergal, taking the $x$-derivative, using the periodicity of the integrand in imaginary direction, choosing as a proper closed contour an infinitely long rectangle of height $i\pi$, applying Cauchy's residue theorem, integrating with respect to $x$, identifying the correct integration constant by looking at the asymptotics for large imaginary $x$, and finally sending $x$ to zero), we find the following explicit expression,
\begin{equation}\label{eq:alpha_gapless_odd_roots_of_unity}
  \alpha(\Delta_m) \;=\; \ln(2) \,+\, \frac{1}{2m+1}\sum_{j=1}^{2m} \frac{\sin^2\!\left(\frac{\pi j}{2m+1}\right)}{\cos\mspace{-2mu}\left(\frac{2 \pi j}{2m+1}\right)}\left[\psi\mspace{-2mu}\left(\frac{j}{2m+1}\right)+3\ln(2)+\gamma_E\right]
\end{equation}
with the digamma function $\psi$ and Euler's constant $\gamma_E$. Simple examples of Eq.~\eqref{eq:alpha_gapless_odd_roots_of_unity} are $m=0,1,2$: $\alpha(-1)=\ln(2)$, $\alpha(1/2)=\ln(3\sqrt{3}/4)$, and $\alpha(\cos(\pi/5))=\ln[4/(1+\sqrt{5})]$. Further explicit expressions are $\alpha(0)=\ln(\sqrt{2})$ and $\alpha(1/\sqrt{2})=3\ln(2)/4 - G/\pi$ with Catalan's constant $G$. 

The decay rate at the isotropic point $\Delta=1$ can be computed by taking the limit $\gamma\to 0$ in Eq.~\eqref{eq:alpha_planar_b}. The result reads 
\begin{equation}\label{eq:alpha_isotropic}
  \alpha(1) \;=\; \ln\left[2\sqrt{\pi}\,\frac{\Gamma(\frac{3}{4})}{\Gamma(\frac{1}{4})}\right] \;=\; 0.18077055\ldots\epp
\end{equation}

Let us now consider the axial regime. The positive ground state Bethe roots are bounded, $0 < \lambda_j < \pi/2$, and $\rho$ is supported on the interval $\lambda\in [-\pi/2,\pi/2]$ (see e.g.~Eq.~\eqref{eq:Bethe_roots_scaling} and Fig.~\ref{fig:rhos}). Thus, in Eq.~\eqref{eq:alpha_via_logP}, we now have to integrate from zero to $\pi/2$ rather than from zero to infinity as in the planar and isotropic cases. The decay rate can be simplified to ($\Delta=\cosh\eta$)
\begin{subequations}\label{eq:alpha_axial}
\begin{align}
  \alpha(\Delta) &\;=\; -\int_0^{\pi/2} \ln\left[\frac{\tan(\lambda+\frac{i\eta}{2})\tan(\lambda-\frac{i\eta}{2})}{4\sin^2(2\lambda)}\right]\rho(\lambda)\: {\rm d}\lambda \\[1.ex]
  &\;=\; -\sum_{k=1}^\infty\, \frac{1 - e^{-\eta k} + (-1)^k(1 + e^{-k\eta})}{2k\cosh(\eta k)} \epp
\end{align}
\end{subequations}
It vanishes in the Ising limit $\Delta\to\infty$, $\lim_{\Delta\to\infty} \alpha(\Delta)=0$, as expected (see text below Eq.~\eqref{eq:OL_result_axial} in Sec.~\ref{sec:introduction}). 

\begin{figure}
\begin{center}
  \includegraphics[width=0.52\columnwidth]{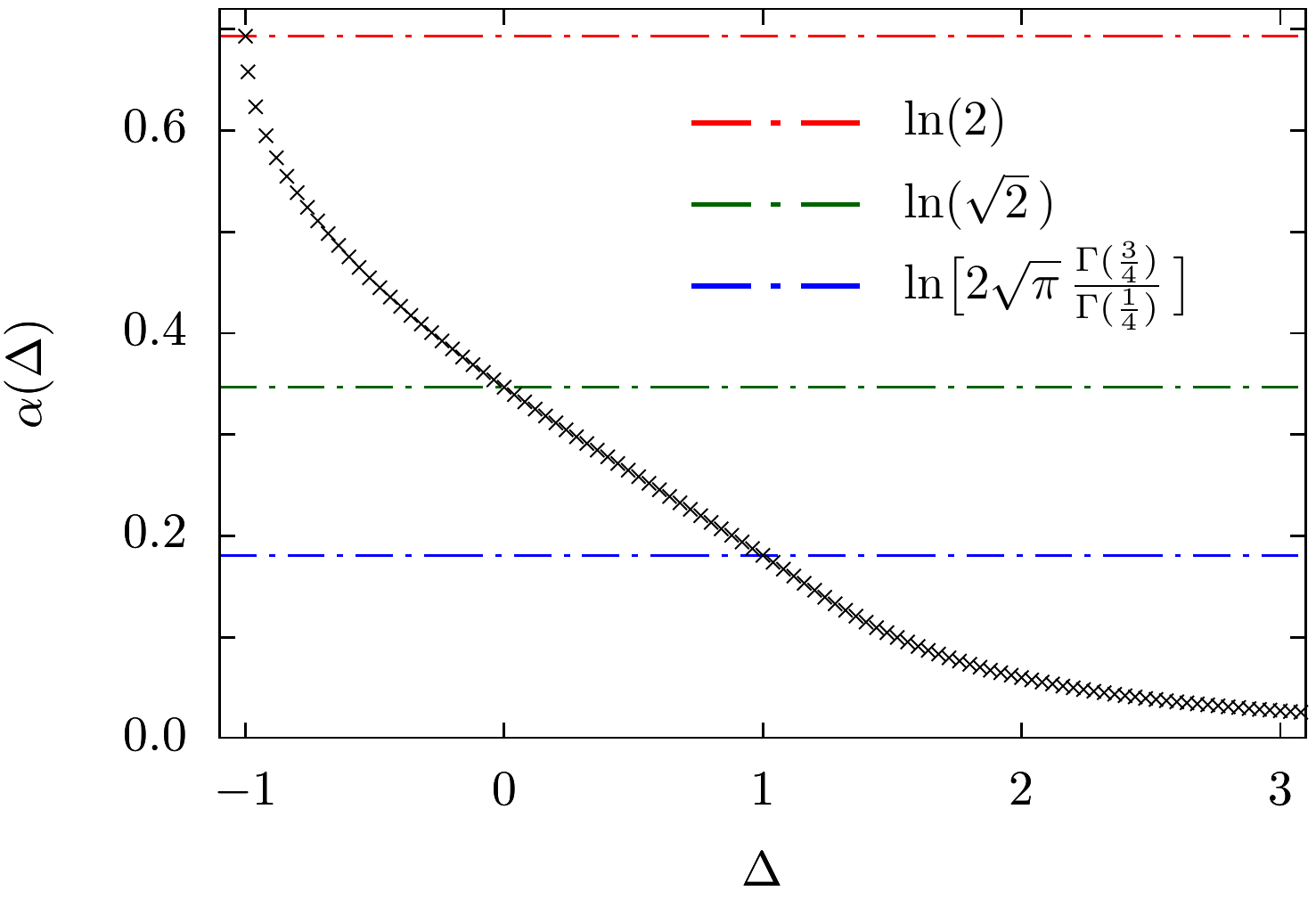}
  \caption{Decay rate $\alpha$ as a function of $\Delta$. It is smooth at the phase-separating point $\Delta=1$. Horizontal lines indicate the values of $\alpha$ at $\Delta=-1$, $0$, $1$ (see text).}
  \label{fig:alpha_of_Delta}
\end{center} 
\end{figure}

The curve $\alpha(\Delta)$ is shown in Fig.~\ref{fig:alpha_of_Delta}, together with the explicit values $\alpha(-1)$, $\alpha(0)$, and $\alpha(1)$. It is a smooth function in the whole regime $\Delta > -1$, even at the point $\Delta=1$. 

\section{Approximating sums over ground state Bethe roots by means of the Euler-Maclaurin formula}\label{app:EMF}

In this section we study how sums over the positive ground state Bethe roots $\{\lambda_j\}_{j=1}^{L/4}$ with $0<\lambda_1<\ldots<\lambda_{L/4}$ can be approximated by means of the Euler-Maclaurin formula (see e.g.~Ref.~\cite{Woynarovich89}). In particular, we are interested in sums like $\sum_{j=1}^{L/4} f(\lambda_j)$ where $f$ is a function with logarithmic singularities at the endpoints, as e.g.~in Eq.~\eqref{eq:P(L)}. 

The two endpoints are given by the smallest and largest positive ground state Bethe roots, $\lambda_1$ and $\lambda_{L/4}$, respectively. Therefore, we have to analyze how these two Bethe roots scale in the TDL. To this end, we compare integrals of the discrete ground state distribution $\rho_L$, Eq.~\eqref{eq:rhoL_discrete}, with integrals of the corresponding TDL distribution $\rho$, Eq.~\eqref{eq:rho_infty}, both over the same interval. In the planar regime, for instance, the primitive functions of $\rho_L$ and $\rho$ respectively read
\begin{align}
  \sigma_L(\lambda) &\;=\; \int_{-\infty}^\lambda\rho_L(\lambda'){\rm d}\lambda' \;=\; \frac{1}{L}\sum_{j=1}^{L/4}\left[\Theta(\lambda-\lambda_j) + \Theta(\lambda+\lambda_j)\right]\epc \\[0.2ex]
  \sigma(\lambda)   &\;=\; \int_{-\infty}^\lambda\rho(\lambda'){\rm d}\lambda'\;\; \;=\; \frac{1}{4}\,+\,\frac{1}{\pi}\arctan\!\left[\tanh\!\left(\frac{\pi \lambda}{2\gamma}\right)\right]\epc
\end{align}
where $\Theta$ is the Heavyside step function, i.e.~$\Theta(x)=0$ for $x<0$, $\Theta(0)=1/2$, and $\Theta(x)=1$ for $x>0$. Therefore, $\sigma_L(\lambda_1)-\sigma_L(0)=1/(2L)$. Identifying this with $\sigma(\lambda_1)-\sigma(0)$, we obtain $\lambda_1 = \frac{2\gamma}{\pi}{\rm artanh}[\tan(\frac{\pi}{2L})]$ which implies $\lambda_1 \simeq \gamma/L$ in the limit of large system size $L$. Similarly, in order to compute the large system size scaling of $\lambda_{L/4}$, we identify $\sigma_L(\infty)-\sigma_L(\lambda_{L/4}) = 1/(2L)$ with 
\begin{equation}
  \sigma(\infty)\,-\,\sigma(\lambda_{L/4}) \;=\; \frac{1}{4} \,-\, \frac{1}{\pi}\arctan\!\left[\tanh\!\left(\frac{\pi}{2\gamma}\lambda_{L/4}\right)\right] \;\simeq\; \frac{1}{\pi}e^{-\frac{\pi}{\gamma}\lambda_{L/4}}\epc
\end{equation} 
where we used in the second step that $e^{-\frac{\pi}{\gamma}\lambda_{L/4}}$ is small. Thus, we obtain $\lambda_{L/4} \simeq \frac{\gamma}{\pi}\ln(\frac{2L}{\pi})$. The isotropic limit $\gamma\to 0$ (after scaling all Bethe roots by $\gamma$) can be performed easily, yielding $\lambda_1 \simeq 1/L$ and $\lambda_{L/4} \simeq \frac{1}{\pi}\ln(\frac{2L}{\pi})$. The analysis of the scaling in the axial regime is not much more involved and proceeds similarly. Without presenting the details of this calculation, we refer to the final result, Eq.~\eqref{eq:Bethe_roots_scaling}. 

Let us now consider a sum $\sum_{j=1}^{L/4}f(\lambda_j)$ over the positive ground state Bethe roots $\{\lambda_{j}\}_{j=1}^{L/4}$. We are interested in the large system size scaling of this sum. In particular, we would like to compute the extensive part, which is proportional to $L$, as well as the order one contribution by means of the Euler-Maclaurin formula (EMF). Denoting the smallest and largest positive ground state Bethe roots by $\epsilon_L = \lambda_1$ and $\Lambda_L = \lambda_{L/4}$, respectively, a slightly modified version of the EMF reads \cite{Woynarovich89}
\begin{equation}\label{eq:EMF}
  \sum_{j=1}^{L/4}f(\lambda_j) \simeq L\int_{\epsilon_L}^{\Lambda_L} \!f(\lambda)\rho(\lambda)\,{\rm d}\lambda \,+\, \frac{f(\Lambda_L)+f(\epsilon_L)}{2} \,+\, \sum_{n=1}^\infty \frac{B_{2n}}{(2n)!} \left(\frac{1}{L\rho(\lambda)}\frac{\rm d}{{\rm d}\lambda}\right)^{2n-1}\!f(\lambda)\;\Big|^{\Lambda_L}_{\epsilon_L}\epp\quad
\end{equation}
Here, $B_{2n}$ are the Bernoulli numbers and $g(\lambda)\big|^{\Lambda}_{\epsilon} = g(\Lambda)-g(\epsilon)$. The main modification as compared to the original EMF (see e.g.~Ref.~\cite{Costin08}) is that we replaced the measure ${\rm d}j$ under the integral by $L\rho(\lambda){\rm d}\lambda$ as well as the derivative $\frac{\rm d}{{\rm d}j}$ by $\frac{1}{L\rho(\lambda)}\frac{\rm d}{{\rm d}\lambda}$. 

In the following analysis, we use for $\rho$ the TDL ground state distribution of Eq.~\eqref{eq:rho_infty}. This is of course a delicate point \cite{Kozlowskicondensation} since we completely neglect finite-size corrections to the root distribution $\rho$. The above replacements are fine for functions that are regular in the interval $[\epsilon_L, \Lambda_L]$ (see Ref.~\cite{Kozlowskicondensation} for more details), in particular at the endpoints $\lim_{L\to\infty}\epsilon_L = 0$ and $\lim_{L\to\infty}\Lambda_L = \Lambda_\infty$. Here, regular means that $f$ is bounded and almost constant in the intervals $[0,\epsilon_L]$ and $[\Lambda_L,\Lambda_\infty]$. Then, the sum in the second line of Eq.~\eqref{eq:EMF} is subleading. Further, the integral in Eq.~\eqref{eq:EMF} can be simplified in the following way, using that $\int_{0}^{\epsilon_L}\rho(\lambda)\,{\rm d}\lambda = \int_{\Lambda_L}^{\Lambda_\infty}\rho(\lambda)\,{\rm d}\lambda = 1/(2L)$,
\begin{equation}\label{eq:integral_EMF_regular}
  L\int_{\epsilon_L}^{\Lambda_L}f(\lambda)\rho(\lambda){\rm d}\lambda \;\approx\; L\int_{0}^{\Lambda_\infty}f(\lambda)\rho(\lambda){\rm d}\lambda \;-\; \frac{f(\epsilon_L)+f(\Lambda_L)}{2}\epp
\end{equation}
Hence, the exptensive part of the left-hand side of Eq.~\eqref{eq:EMF} is given by the first term, $L\int_{0}^{\Lambda_\infty}f(\lambda)\rho(\lambda){\rm d}\lambda$, and there is in total no order one contribution. We checked this for several examples of functions $f$ by solving Bethe equations and computing the sum of the left-hand side of Eq.~\eqref{eq:EMF} explicitly. 

On the contrary, for a function $f$ that has logarithmic singularities at the endpoints, e.g.~$f=\ln(p)$, the analysis is more involved. Let us first focus on the lower boundary, since the calculation is basically the same in all three cases `planar', `isotropic', and `axial'. Let us assume that the asymptotics of $f$ for small positive arguments is given by $f(\lambda)\simeq -2\ln(\lambda) + \mathcal{O}(\lambda)$, which is fulfilled for $f=\ln(p)$. Then, we find
\begin{subequations}
\begin{align}\label{eq:EMF_log_at_zero}
  L\int_{\epsilon_L}^{\Lambda_L}\!f(\lambda)\rho(\lambda)\:{\rm d}\lambda &\;\approx \; L\int_{0}^{\Lambda_L}\!f(\lambda)\rho(\lambda)\:{\rm d}\lambda \,+\, \ln(\epsilon_L)\,-\,1\epc\\[1.2ex]
  \sum_{n=1}^\infty \frac{B_{2n}}{(2n)!} \left(\frac{1}{L\rho(\lambda)}\frac{\rm d}{{\rm d}\lambda}\right)^{2n-1}\!f(\lambda)\Big|^{\Lambda_L}_{\epsilon_L} &\;\simeq\; (\text{$\Lambda_L$-term}) \,+\,\sum_{n=1}^\infty \frac{B_{2n}}{n(2n-1)} \left(\frac{1}{L\rho(\epsilon_L)\epsilon_L}\right)^{2n-1}\notag\\[1.2ex]
  &\;=\; (\text{$\Lambda_L$-term})\,-\,\ln(2)\,+\,1\epc\label{eq:EMF_log_at_zero2}
\end{align}
where `$\Lambda_L$-term' denotes the contribution of the upper endpoint on the left-hand side of Eq.~\eqref{eq:EMF_log_at_zero2}, see below.
\end{subequations}
In the very last step, we used the expansion of the log-gamma function (see e.g.~Ref.~\cite{Pearce07})
\begin{equation}\label{eq:log_gamma}
  2\ln[\Gamma(x+1)] \;=\; 2x(\ln(x)-1) \,+\, \ln(x) \,+\, \ln(2\pi) \,+\, \sum_{n=1}^\infty \frac{B_{2n}}{n(2n-1)} \frac{1}{x^{2n-1}}\epc
\end{equation} 
with argument $x=L\rho(\epsilon_L)\epsilon_L \simeq 1/2$. Therefore, the contribution to the order one term of the sum $\sum_{j=1}^{L/4}\ln[p(\lambda_j)]$ from the lower endpoint is in total\ \;$-\ln(2)$. The same line of arguments holds for the upper endpoint in the axial regime, $\Lambda_L \simeq \pi/2 - \pi\eta/(2kK\mspace{-1mu}(k)L)$, see Eq.~\eqref{eq:Bethe_roots_scaling}, yielding as a second contribution again\ \;$-\ln(2)$. This confirms the order one term \eqref{eq:P_0_result_axial}, which we also rigorously derived in Sec.~\ref{sec:Ps_order_one_contribution} by a completely independent approach (NLIE). The upper endpoint in the two other cases (planar and isotropic), $\Lambda_\infty = \infty$, causes serious problems. The main obstacle is that the finite-size corrections of $\rho$, which are neglected in Eq.~\eqref{eq:EMF}, become important for functions that scale for \textit{large} arguments $\lambda$ as\ \;$\sim \ln(\lambda)$. Again, we have checked this by solving Bethe equations for finite, but large system sizes up to $L=8192$. Computing the finite-size corrections of $\rho$ is a difficult task. 

In conclusion, the EMF approach to compute the sum $\sum_{j=1}^{L/4}\ln[p(\lambda_j)]$ works well in the axial regime but, unfortunately, leads to unsurmountable difficulties in the planar regime and at the isotropic point.

\section*{References}

\bibliographystyle{iopart-num}
\bibliography{overlap_literature}

\end{document}